\PassOptionsToPackage{pagewise}{lineno}
\documentclass{iopjournal}
\usepackage[square, numbers]{natbib}
\usepackage{hyperref}
\usepackage{booktabs}
\usepackage{lmodern}
\usepackage{microtype}
\usepackage{amsmath}
\usepackage{amssymb}
\usepackage{bookmark}
\usepackage{cleveref}
\usepackage{lineno}

\begin{document}

\articletype{Paper}    
\\
\title{Drift wave turbulence and transport analysis of hot ion L-mode plasma with fishbone instability on HL-3 tokamak}

\author{Y.K. Shu$^{1}$\orcid{0009-0007-4551-6253}, W. Chen$^{1,*}$\orcid{0000-0002-9382-6295},  Z.J. Li$^{2}$\orcid{0009-0006-1394-8134}, S. Xu$^{1}$\orcid{0000-0000-0000-0000}, Y.F. Liu$^{1}$\orcid{0009-0000-0000-0000}, B.L. Hao$^{1}$\orcid{0000-0003-3009-0660},  Z. Wang$^{1}$\orcid{0009-0000-5710-0318}, D.L. Yu$^{1}$\orcid{0000-0002-5503-7957}, S.S. Yang$^{1}$\orcid{0000-0000-0000-0000}, W.Y. Zhai$^{1}$\orcid{0000-0002-5477-7339}, X.L. Zhu$^{1}$\orcid{0000-0001-7467-5287}, P.W. Shi$^{1}$\orcid{0000-0002-2071-7741}, X.L. Xu$^{1}$\orcid{0000-0002-0000-0000}, T. Wu$^{1}$\orcid{0000-0003-2269-1968}, Z.B. Shi$^{1}$\orcid{0000-0002-7640-9844} and HL-3 Team$^{a}$}

\affil{$^{1}$Center for Fusion Science, Southwestern Institute of Physics, P.O. Box 432 Chengdu 610041, China}

\affil{$^{2}$Faculty of Arts and Sciences, Beijing Normal University, Zhuhai 519087, China}

\affil{$^*$Author to whom any correspondence should be addressed.}

\affil{$^a$See Duan et al 2024 (\url{https://doi.org/10.1088/1741-4326/ad6e9e}) for the
HL-3 Team.}

\email{chenw@swip.ac.cn}


\begin{abstract}
An integrated analysis of the HL-3 hot-ion L-mode (shot \#11972, $(T_i(0)/T_e(0))_{\max}\approx3.5$, $T_i(0)\approx9.8$\,keV) traces how a single $m=1,n=1$ fishbone couples toroidal rotation to ion-scale drift-wave turbulence. A steep fast-ion pressure gradient under 2.5\,MW NBI excites the fishbone, which, synchronized with the sawtooth cycle, redistributes fast ions, while the sawtooth crash collapses rotation from above $100$ to $60$\,krad/s. OMFIT modeling with TGLF shows the fast ions stabilize ITG/TEM turbulence and hold $q_{\min}\approx 1$, favoring the ion internal transport barrier. The crash temporarily weakens the stabilization from the high temperature ratio, fast-ion dilution, and $E\times B$ shear. Flux and eigenvalue scans confirm the temperature ratio and fast ions suppress transport together. As $T_i/T_e$ rises, the high wavenumber ($k_y\rho_s > 1$) drift waves retreat from the ion to the electron diamagnetic direction, an ion temperature screening that stabilizes the ion-direction ITG. Steepening $a/L_{n_e}$ instead dilutes the temperature-gradient drive through $\eta = L_n/L_T$, quenching the high-wavenumber branches and turning them from the electron to the ion direction. At $T_i/T_e = 4.77$ (three times the initial value) the ITG growth rate is strongly suppressed, even without fast ions, inside the window $8 < a/L_{T_i} < 12$. These results trace a cross-scale coupling loop spanning MHD, fast-ion transport, and microturbulence relevant to fast-ion confinement and rotation sustainment in the deuterium-tritium phase.
\end{abstract}

\section{Introduction}

The hot ion mode (HIM), in which the ions are hotter than the electrons ($T_i/T_e > 1$), is a leading operating scenario for next-generation reactors~\cite{Kamada2002, Han2022}. Hotter ions raise the fusion reaction rate, and pairing them with a peaked density and a quiescent core improves both the triple product $nT\tau_E$ and the gain $Q$~\cite{Gormezano1998, Kishimoto2005}. Internal transport barriers (ITBs) with ion temperatures near $45\,\mathrm{keV}$ were achieved on JET and JT-60U~\cite{Gormezano1998, Kishimoto2005, de2009internal}, and ITB H-modes at cooler ion temperatures followed on DIII-D and ASDEX Upgrade~\cite{Lazarus2006, Gruber1999}. KSTAR and ST40 later carried the scenario to long-pulse and compact machines~\cite{Han2022, McNamara2023}. A recurrent theme across these devices is what sustains the ITB: reversed magnetic shear on the large tokamaks~\cite{de2009internal, fujita1999high}, versus rotation-driven $E\times B$ shear on the compact, strongly rotating ones, with fast ions damping the turbulence in both cases. On HL-2A, integrated OMFIT modeling~\cite{meneghini2015integrated} of high-$\beta_N$ discharges with two transport barriers traced the core turbulence suppression to fast ions together with $E\times B$ shear~\cite{Li2024}. The same workflow later addressed core-compatible H-mode divertor detachment~\cite{shu2025integrated}. HL-3 recently exceeded $10\,\mathrm{keV}$ in ion temperature ($T_i/T_e \approx 2.4$)~\cite{shu2026integrated} and reached a triple product of $0.65\times 10^{20}\,\mathrm{m}^{-3}\cdot\mathrm{keV}\cdot\mathrm{s}$~\cite{zhong2025breakthrough}, a national benchmark for the planned deuterium-tritium campaign~\cite{deuterium-tritium2026}. These studies still center on producing and characterizing the hot ion state. None has traced how one macroscopic fast-ion-driven MHD mode, the fishbone, links the momentum balance to microturbulence stability. The present work fills that gap.

Ion-scale turbulence, dominated by the ion temperature gradient (ITG) mode, sets the anomalous transport in the tokamak core~\cite{horton1999drift}. Fast ions damp this turbulence and improve confinement~\cite{na2025fast}. How fast ions modify turbulent transport is an active question at the interface of these two fields. In the HIM the ion-scale turbulence is stabilized jointly by the elevated temperature ratio \(T_{\mathrm{i}}/T_{\mathrm{e}}\), the fast-ion dilution of the thermal-ion drive, and the \(E\times B\) shear from the co-injected NBI torque. The temperature ratio raises the ITG excitation threshold and acts as an ion temperature screening, damping the ion-directed drift waves at short wavelength and pushing the high-wavenumber turbulence toward the electron direction, so the ion-direction ITG is self-consistently stabilized. Fast-ion dilution raises the threshold further~\cite{jung2026fast}, and the shear contributes a rate-based quenching channel. Their combined action gives the HIM its intrinsic transport advantage and frames the integrated analysis that follows.
In high-power neutral beam injection (NBI) heated plasmas, energetic particle-driven instabilities decide the confinement~\cite{fasoli2006progress}. Among these, the fishbone mode is particularly important~\cite{heidbrink2011characterization}: an $m=1$, $n=1$ internal kink that trapped fast ions resonantly excite~\cite{chen1984excitation, coppi1986fishbone}, with a frequency that chirps downward, convex during the initial burst and concave as saturation approaches~\cite{berk1996nonlinear}. The chirp encodes the phase-space dynamics of fast ions and the onset of $E\times B$ velocity-shear damping~\cite{fu2006nonlinear}, and sets the nonlinear saturation amplitude and its impact on fast ion confinement~\cite{fasoli2006progress}. The fishbone mode also helps ITB formation: the $m=1,n=1$ resonance requires $q_{\min}\approx 1$, and the mode holds $q_{\min}$ near unity, sustaining a weak (near-zero) magnetic shear configuration favorable for ion internal transport barrier (iITB) establishment~\cite{chen2010features, ge2023multiple}.

Toroidal rotation couples NBI momentum input to turbulence stabilization~\cite{burrell1999tests, terry2000suppression}. Co-injected NBI beams drive core toroidal rotation above $100\,\mathrm{krad/s}$ in HL-3 discharges. The rotation gradient generates a radial electric field $E_r$. Its $E\times B$ flow shearing rate $\gamma_{E\times B}$ tears apart turbulent eddies and suppresses ion temperature gradient (ITG) and trapped electron mode (TEM) transport~\cite{biglari1990influence, garbet1990magnetic}, a mechanism central to transport-barrier formation~\cite{hahm1995flow, itoh1996role}. On HL-3, locked tearing modes can likewise induce a strong rotation shear that suppresses turbulence and forms an ITB with $T_i > 9\,\mathrm{keV}$~\cite{yan2026lift}. The fast ions injected by NBI act through several channels. They deposit the toroidal momentum that drives this rotation and its $E\times B$ shear. Through dilution and wave-particle interaction, they raise the drift-wave threshold and stabilize the modes~\cite{Han2022}. The fast-ion pressure gradient can further sustain a fast-ion-induced internal transport barrier (F-ATB)~\cite{di2021new}. The spatial distribution of the fast ions that carry the NBI torque is critical~\cite{pankin2004tokamak}. When fast ion redistribution decouples the torque source from the plasma response, rotation collapse can cause ITB loss~\cite{de2009internal}.
Fast-ion transport and its interplay with magnetohydrodynamic (MHD) activity govern HIM dynamics~\cite{fasoli2006progress, heidbrink2011characterization}. During fishbone bursts, fast ions spread radially in an avalanche-like way, releasing the steep core fast ion pressure gradient~\cite{chen1984excitation, fu2006nonlinear}. The transport is nonlinear: the mode excites phase-space structures (holes and clumps) that resonantly scatter fast ions, and the resulting gradient relaxation feeds back on the mode drive~\cite{berk1996nonlinear, fu2006nonlinear}.

The cross-scale coupling between macroscopic MHD and microscopic turbulence is the central theme of this work. At the macroscopic scale, the $m=1$, $n=1$ fishbone mode drives global fast ion transport and cyclic sawtooth crash dynamics. During ITB rebuilding, the safety factor minimum $q_{\min}$ oscillates around unity~\cite{chen1984excitation, vongoeler1974studies}. At the microscopic scale, drift-wave instabilities (ITG, TEM, and ETG modes) set turbulent transport levels across the plasma cross section~\cite{waltz1997gyro, staebler2005gyro}. The fishbone mode bridges these two scales. Synchronized with the sawtooth cycle, it governs the radial distribution of resonant fast ions. The sawtooth crash temporarily decouples the NBI torque source from the core plasma and depletes the stabilizing fast-ion population, while the fishbone itself preserves the fast-ion stabilization that favors the ion internal transport barrier. Macroscopic MHD dynamics couple to microturbulence stability. The pre-existing $n=1$ tearing mode and coexisting $n=1$ kink mode accompany the fishbone mode in the Mirnov spectrum and play a secondary role.

These multi-scale processes are unraveled by integrated modeling: OMFIT~\cite{meneghini2015integrated} couples the EFIT equilibrium reconstruction~\cite{lao1985separation}, the NUBEAM and ONETWO source solvers~\cite{pankin2004tokamak, stjohn1994transport}, and the TGYRO transport solver, whose TGLF and NEO modules supply the turbulent and neoclassical fluxes~\cite{waltz1997gyro, staebler2005gyro, pan2017investigation, belli2008kinetic}. A companion study~\cite{shu2026integrated} runs the same machinery to characterize turbulent transport under strong \(E\times B\) shear in the HL-3 HIM, including the fishbone mode and toroidal rotation. The present work shares that machinery but follows a distinct physics path, opened up by the high fast-ion beta of shot \#11972. It extends the analysis to the momentum channel and isolates the fast-ion contribution by repeating each instability and flux scan with the fast-ion population removed, to trace the cross-scale coupling through which a single fishbone mode couples both momentum balance and microturbulence stability.

In this paper, we present an integrated experimental and simulation study of the cross-scale evolution of fishbone modes and their coupling to toroidal rotation and NBI torque deposition in the HL-3 HIM. Section~2 reports the experimental fishbone observations, Sections~3 and~4 the kinetic-profile reconstruction and the deposition, beta, and transport results, Section~5 the TGLF-scan microturbulence analysis, and Section~6 the principal findings and their broader implications.

\section{Hot ion mode scenarios with fisbone instabilities}
\begin{figure}[htbp]
 \centering
        \includegraphics[width=1.0\textwidth]{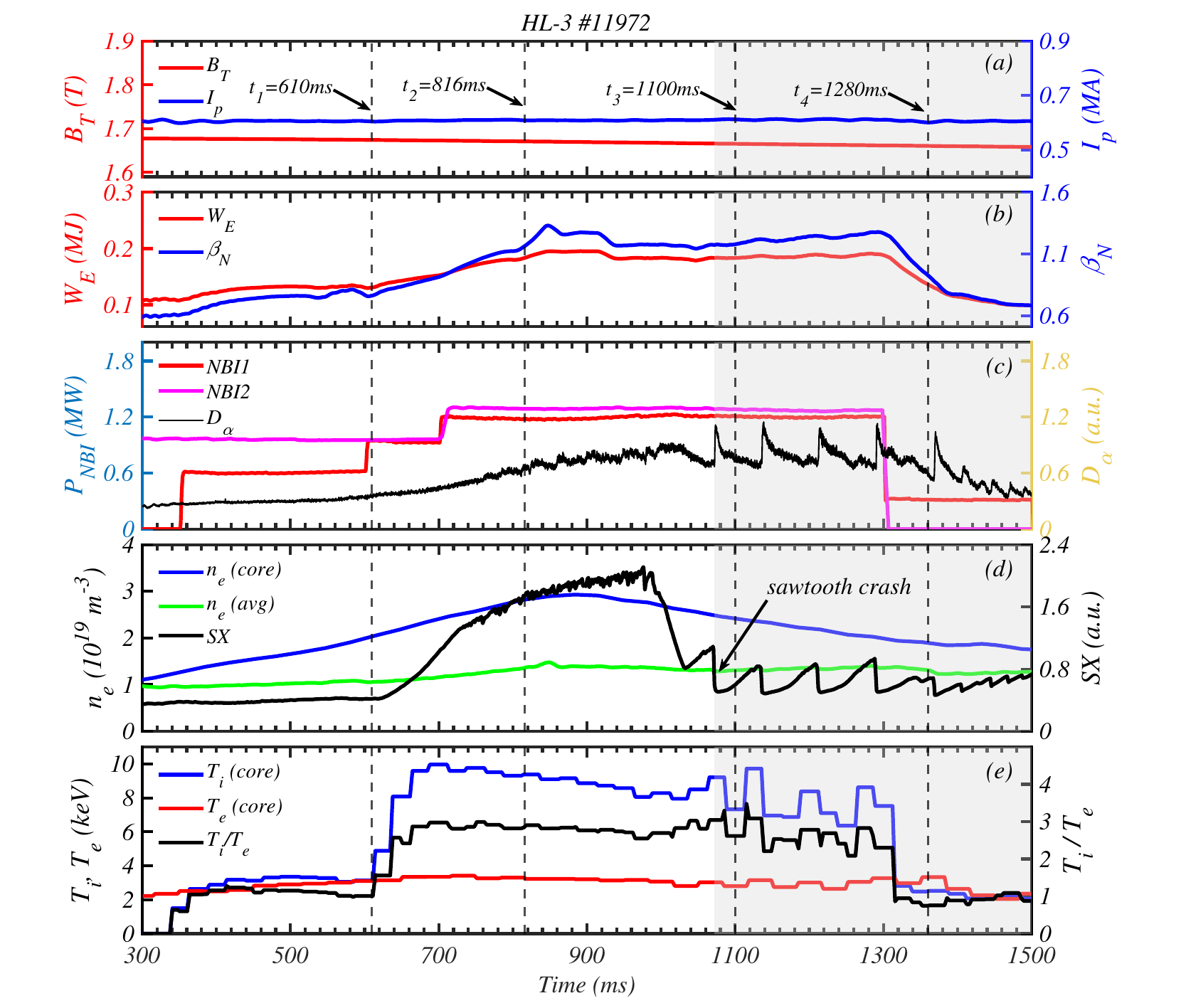}
 \caption{Key parameter evolution for HL-3 shot \#11972. (a) plasma current and toroidal magnetic field; (b) stored energy and normalized beta $\beta_{\mathrm{N}}$; (c) NBI heating power of the two co-injected beams and $D_\alpha$ emission; (d) line-averaged electron density, soft X-ray signal, and core electron density; (e) core ion temperature, core electron temperature, and $T_i/T_e$ ratio. The four dashed lines mark the selected simulation time slices $t_1 = 610\,\mathrm{ms}$, $t_2 = 816\,\mathrm{ms}$, $t_3 = 1100\,\mathrm{ms}$, and $t_4 = 1280\,\mathrm{ms}$.}
\label{fig1}
\end{figure}

In the companion discharge \#13243, recent HL-3 campaigns have already exceeded a central ion temperature of $10\,\mathrm{keV}$~\cite{shu2026integrated}, alongside a fusion triple product of $0.65 \times 10^{20}\,\mathrm{m}^{-3}\cdot\mathrm{keV}\cdot\mathrm{s}$, the current HL-3 benchmark.
The present study turns to shot \#11972, a HIM discharge with strong fishbone mode activity.
The plasma current was $0.62\,\mathrm{MA}$ and the toroidal magnetic field was $1.67\,\mathrm{T}$ (Fig.~\ref{fig1}a).
The stored energy peaked at $0.191\,\mathrm{MJ}$, and the normalized beta $\beta_{\mathrm{N}}$ was about $1.3$ (Fig.~\ref{fig1}b).
The auxiliary heating used two co-injected neutral beams~\cite{wagner1982regime} with a combined power of $2.5\,\mathrm{MW}$. The first beam reached $1.2\,\mathrm{MW}$ in three steps, and the second reached $1.3\,\mathrm{MW}$ in two steps (Fig.~\ref{fig1}c).
The line-averaged electron density reached a maximum of $1.4 \times 10^{19}\,\mathrm{m^{-3}}$ (Fig.~\ref{fig1}d).
At about $600\,\mathrm{ms}$, when the NBI power reached $2\,\mathrm{MW}$, fishbone modes began to appear~\cite{chen1984excitation}, and a core iITB formed.
After the NBI power reached its peak, small sawtooth oscillations developed, and the plasma density kept rising.
Around $1080\,\mathrm{ms}$, a large sawtooth crash occurred (Fig.~\ref{fig1}d), together with fluctuations of the core ITB (Fig.~\ref{fig1}e).
Throughout the discharge, the electron temperature showed no significant variation, and the $T_i/T_e$ ratio exceeded $3.5$ at its maximum.

At $600\,\mathrm{ms}$, the rising NBI power enables the core thermal ion profile to develop a steep internal transport barrier (ITB). The co-injected fast ions transfer tangential momentum to the background thermal ions, driving the plasma to toroidal rotation above $100\,\mathrm{krad/s}$ and generating an intense $E\times B$ shearing rate that suppresses the ion temperature gradient (ITG) turbulence~\cite{biglari1990influence, garbet1990magnetic}. This rotation-driven shear sustains the extreme HIM ($T_i \approx 3T_e$, Fig.~\ref{fig1}e) and the high-confinement ITB at $816\,\mathrm{ms}$, and it also disrupts the localized linear eigenmode structure of the $m=1, n=1$ fishbone mode, suppressing its growth rate and keeping the mode from growing without bound. Between $700\,\mathrm{ms}$ and $1000\,\mathrm{ms}$ the fishbone mode reaches a dynamically stable state, keeping a finite saturated amplitude at $816\,\mathrm{ms}$ even with the large fast ion energy.

\begin{figure}[htbp]
 \centering
        \includegraphics[width=1.0\textwidth]{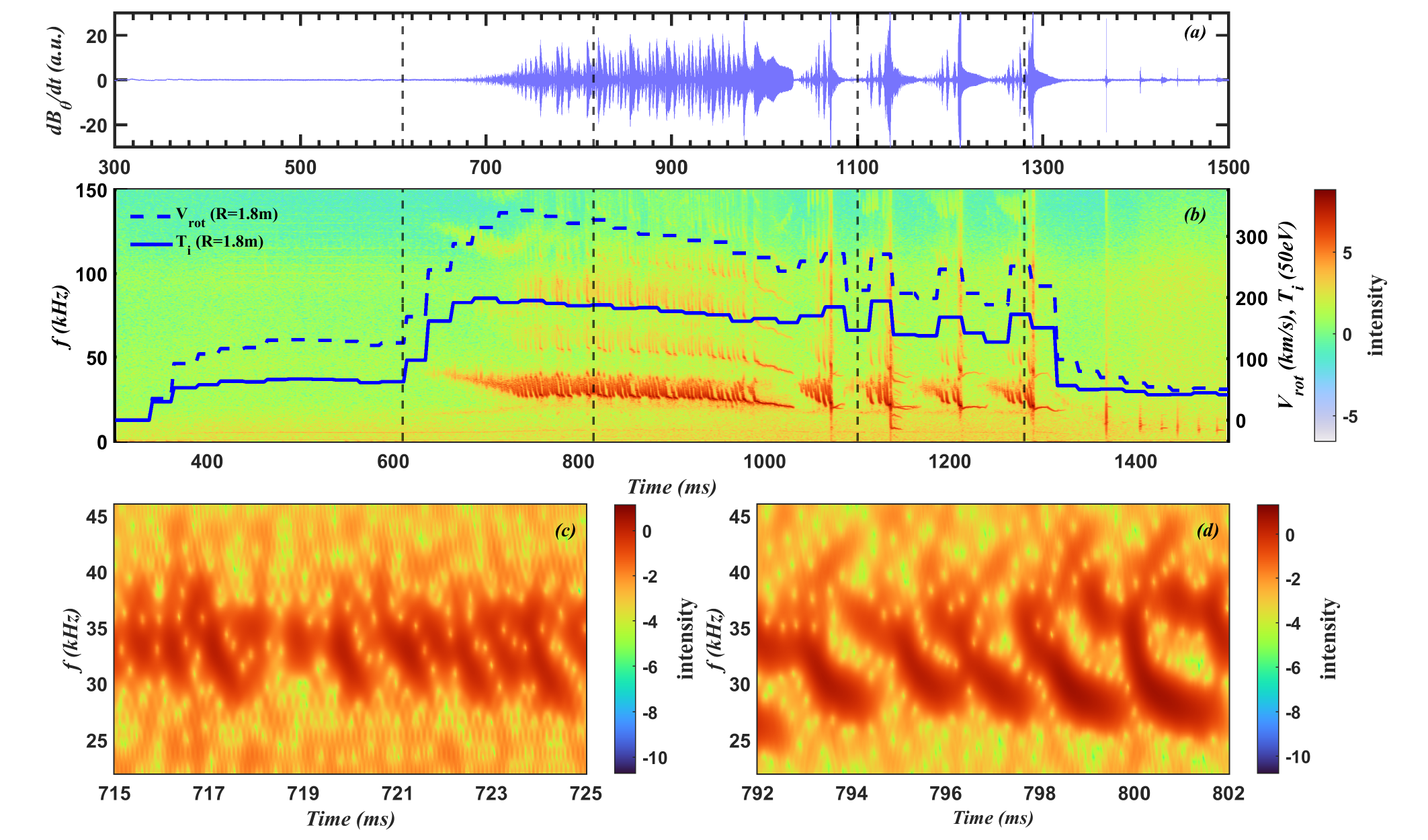}
 \caption{SXR, Mirnov, and CXRS measurements for HL-3 shot \#11972. (a) Mirnov raw signal; (b) Mirnov spectrogram showing the $m=1$, $n=1$ fishbone mode in the $0$--$150\,\mathrm{kHz}$ frequency range, together with an $n=1$ kink mode near $25\,\mathrm{kHz}$ and an $n=1$ tearing mode near $6\,\mathrm{kHz}$, overlaid with ion temperature (blue solid line) and toroidal rotation velocity (blue dashed line) from the CXRS system at $R = 1.8\,\mathrm{m}$; (c) SXR spectrogram in the $715$--$725$\,ms window showing the convex downward chirping; (d) SXR spectrogram in the $792$--$802$\,ms window showing the concave downward chirping. The four dashed lines mark $t_1$--$t_4$.}
\label{fig2}
\end{figure}

A fishbone mode also appears in the L-mode phase of the companion discharge \#13243~\cite{shu2026integrated}, indicating that the fishbone is not an artifact of a single shot. The cross-scale coupling documented here is, however, established in detail for shot \#11972 alone.
The Mirnov probe array spectrogram in figure~\ref{fig2} shows a fishbone mode in the $25$--$45\,\mathrm{kHz}$ frequency gap. Mirnov probe phase measurements determine its poloidal and toroidal mode numbers $m=1$ and $n=1$. The fundamental mode at about $40\,\mathrm{kHz}$ further excites sideband modes at $60\,\mathrm{kHz}$, $80\,\mathrm{kHz}$, and $100\,\mathrm{kHz}$, each with a distinct mode number. The steep fast ion pressure gradient from the large NBI-driven fast ion population drives this mode. The $2.5\,\mathrm{MW}$ of NBI power deposited in the core produces a very high fast ion density and a sharply peaked fast ion pressure gradient. Starting at about $1080\,\mathrm{ms}$, a large sawtooth crash occurs in synchronization with the fishbone mode. Each sawtooth crash and fishbone re-excitation goes with a radial redistribution of core fast ions and outward transport. The core thermal ion temperature and toroidal rotation crash and rebuild together.

\begin{figure}[htbp]
 \centering
        \includegraphics[width=0.9\textwidth]{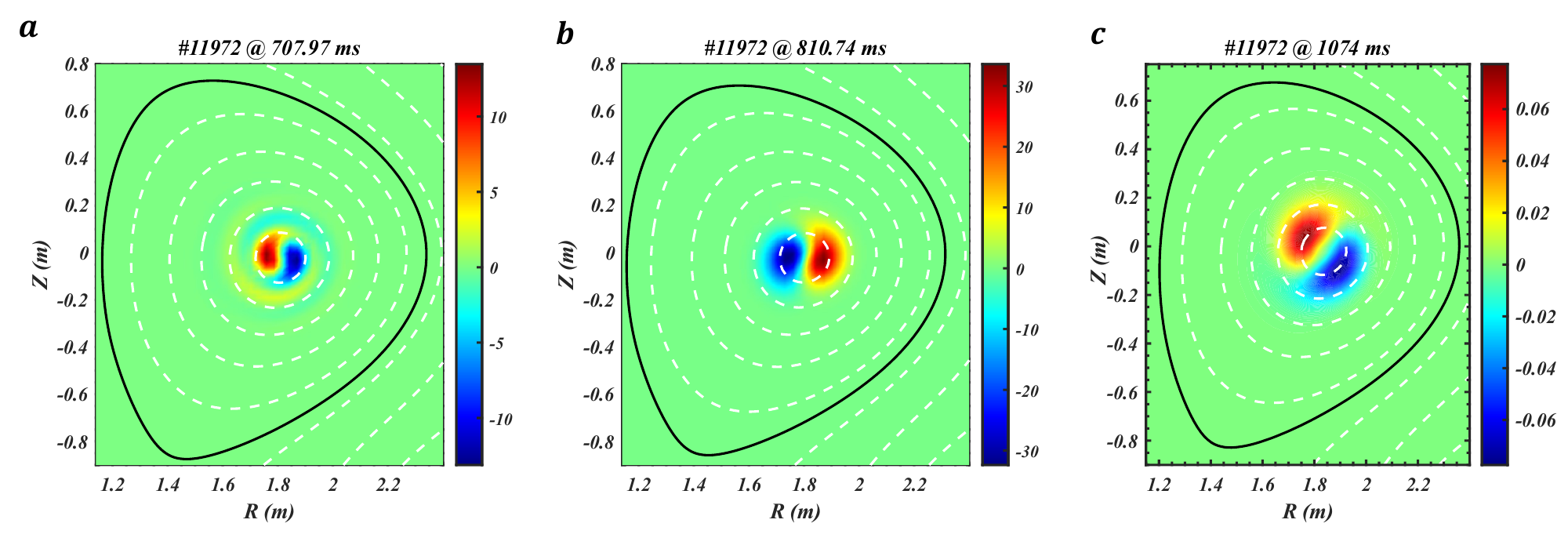}
 \caption{Soft X-ray tomographic inversion of the core plasma redistribution accompanying the fishbone-driven fast-ion transport at three characteristic time windows: (a) $707.97$~ms, (b) $810.74$~ms, and (c) $1074$~ms.}
\label{fig3}
\end{figure}

To further resolve the spatial structure of the fishbone-driven fast ion transport, we performed soft X-ray (SXR) tomographic inversion at three time windows. The convex downward chirping phase during the early nonlinear burst appears at $707.97$~ms (Fig.~\ref{fig3}a). The concave chirping phase approaching nonlinear saturation appears at $810.74$~ms (Fig.~\ref{fig3}b). A third window at $1074$~ms (Fig.~\ref{fig3}c) falls just before the sawtooth crash at $1080$~ms. The SXR emissivity is dominated by bremsstrahlung and recombination radiation from the thermal plasma, and depends on the electron temperature, electron density, and impurity density~\cite{vongoeler1974studies}. The reconstruction therefore tracks the core plasma response to the fishbone-driven fast-ion redistribution.

The inversion results show a clear outward radial movement of the core emissivity as the fishbone mode evolves. At $707.97$~ms, the emissivity concentrates strongly in the core region, consistent with the sharply peaked core fast-ion pressure gradient that drives the initial fishbone burst. By $810.74$~ms, the emissivity peak broadens and shifts outward, reaching its maximum intensity, consistent with the phase-space hole and clump structure resonantly ejecting fast ions as the fishbone mode approaches its nonlinear saturation amplitude. At $1074$~ms, just before the sawtooth crash, the core emissivity drops to its minimum among the three windows, indicating that a large part of the resonant fast ions has moved radially outward and the core fast-ion pressure gradient has largely relaxed, in agreement with the transport modeling of Section~4. This progression, from core peaking through outward expansion to core depletion, gives tomographic evidence, complementary to the modeling, that the fishbone mode drives a continuous radial redistribution of fast ions through its nonlinear saturation cycle, ending in the sawtooth crash.

Figure~\ref{fig2} compares the fishbone mode chirping during two different phases. Panel (c) shows the SXR spectrogram in the $715$--$725$\,ms window, the early nonlinear saturation stage, where the fishbone mode displays a convex downward chirp. In the phase-space hole--clump picture~\cite{berk1996nonlinear,fu2006nonlinear}, this chirp reflects a hole that experiences an intensifying drag as the system enters the strongly nonlinear initial-burst regime, and the high fast-ion energy speeds its outward radial drift toward the resonance boundary. The accompanying avalanche-like ejection of fast ions from the core relaxes the steep fast-ion pressure gradient $\nabla p_{\mathrm{fast}}$.

Panel (d) shows the SXR spectrogram in the $792$--$802$\,ms window, the post-saturation phase, where the chirp lengthens and turns to a concave downward (decelerating) pattern. Within the same hole--clump picture~\cite{berk1996nonlinear,fu2006nonlinear}, the hole now meets strong damping as it moves through the localized $E\times B$ velocity shear, slowing the outward escape of fast ions. The concave chirp thus indicates that the fishbone mode and the fast-ion population have reached a local kinetic equilibrium, with the ambient $E\times B$ shear barrier holding back the ejection rate. In this qualitative picture, the barrier would briefly accumulate fast ions at a localized radial location, an interpretation inferred from the chirp rather than measured directly. This state precedes the macroscopic sawtooth crash at $1080\,\mathrm{ms}$.

\begin{figure}[htbp]
 \centering
        \includegraphics[width=0.9\textwidth]{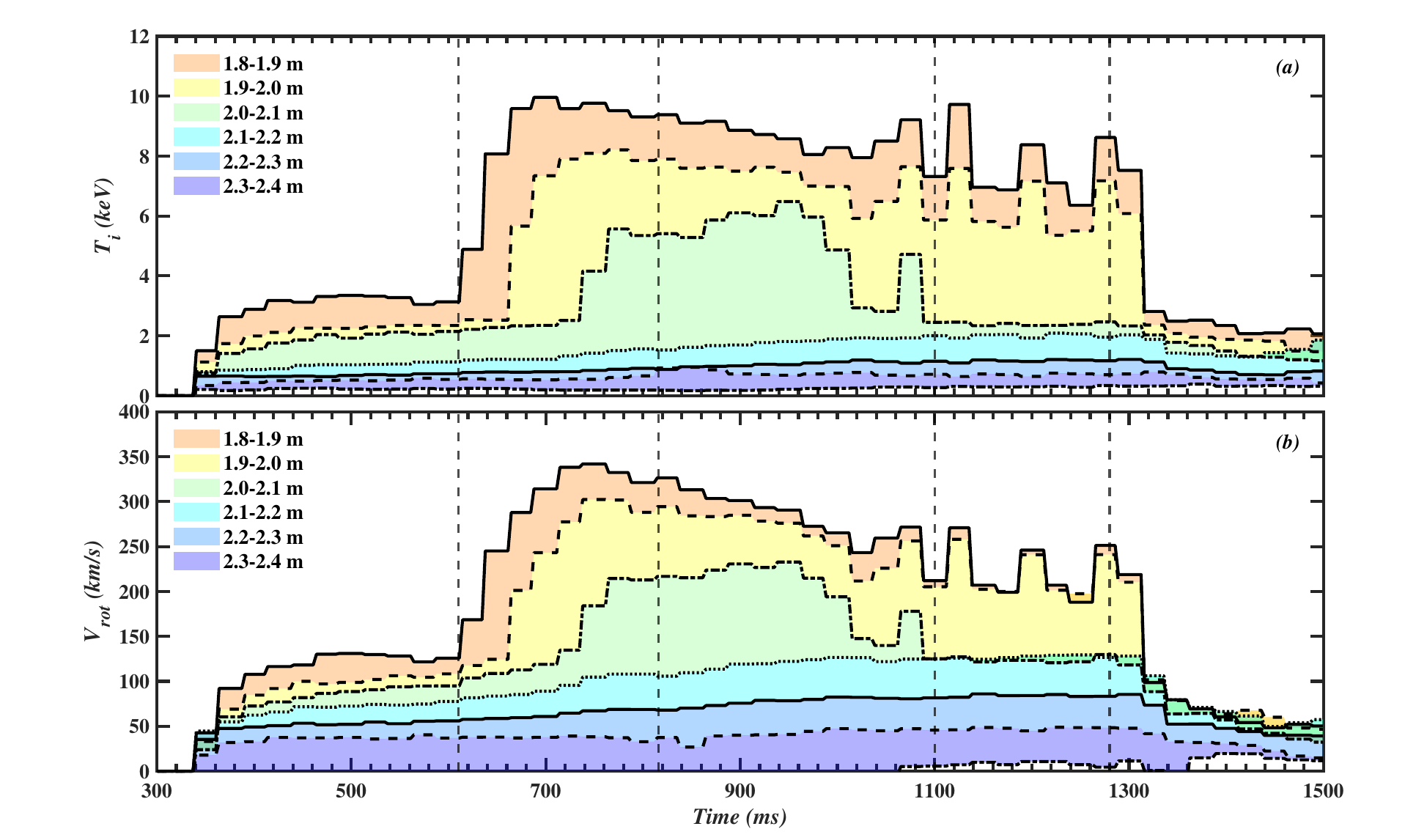}
 \caption{CXRS measurements of (a) ion temperature $T_i$ and (b) toroidal rotation velocity $V_{\phi}$ as functions of major radius $R$ and time for HL-3 shot \#11972. The four dashed lines mark the selected simulation time slices $t_1$--$t_4$.}
\label{fig4}
\end{figure}

The Mirnov coil spectrogram (Fig.~\ref{fig2}b) shows two $n=1$ modes that accompany the fishbone mode (Fig.~\ref{fig2}a). One is an $n=1$ kink mode near $25\,\mathrm{kHz}$ that grows together with the fishbone mode from $620$\,ms. The other is an $n=1$ tearing mode near $6\,\mathrm{kHz}$ in the low-frequency band. It appears as early as $400$\,ms and lasts through the discharge. The kink mode frequency band overlaps with the lower-frequency edge of the fishbone mode. This frequency overlap is consistent with nonlinear coupling between the two modes. The $n=1$ tearing mode, with its much lower frequency, is a long-lived electromagnetic perturbation that exists before the fishbone mode.

The charge exchange recombination spectroscopy (CXRS) diagnostic gives the time evolution of ion temperature and toroidal rotation velocity from the plasma core ($R = 1.8$ to $2.4\,\mathrm{m}$) over $300$--$1500\,\mathrm{ms}$, as shown in figure~\ref{fig4}. During the early NBI phase ($340$--$600\,\mathrm{ms}$), the co-injected fast ion fraction is too small to give much tangential momentum to the background plasma. The toroidal rotation stays low. The core electron density and ion temperature then rise quickly as the beam power reaches its full $2.5$\,MW. The ion temperature profile (Fig.~\ref{fig4}a) develops an iITB structure, and the toroidal rotation (Fig.~\ref{fig4}b) forms a steep shear layer in the core. The NBI power mainly lands in the core, and the edge region ($2.2 < R < 2.4\,\mathrm{m}$) does not develop an ion temperature edge transport barrier. A velocity shear layer still forms there.

In the post-sawtooth phase after $1080\,\mathrm{ms}$, the ion temperature profile keeps a large gradient in the range $1.8 < R < 1.9\,\mathrm{m}$. The toroidal rotation shear layer at this location becomes flat. Both profiles keep their gradients in $1.9 < R < 2.0\,\mathrm{m}$. The gradients of both quantities drop clearly near the ITB foot ($2.0 < R < 2.1\,\mathrm{m}$). This shows that the ITB contracts inward and the parameter values at its foot rise. After $1300\,\mathrm{ms}$, as the NBI power declines, the plasma disrupts and the ITB collapses.

\section{Kinetic profile reconstruction and integrated simulation}

\begin{figure}[htbp]
 \centering
        \includegraphics[width=0.95\textwidth]{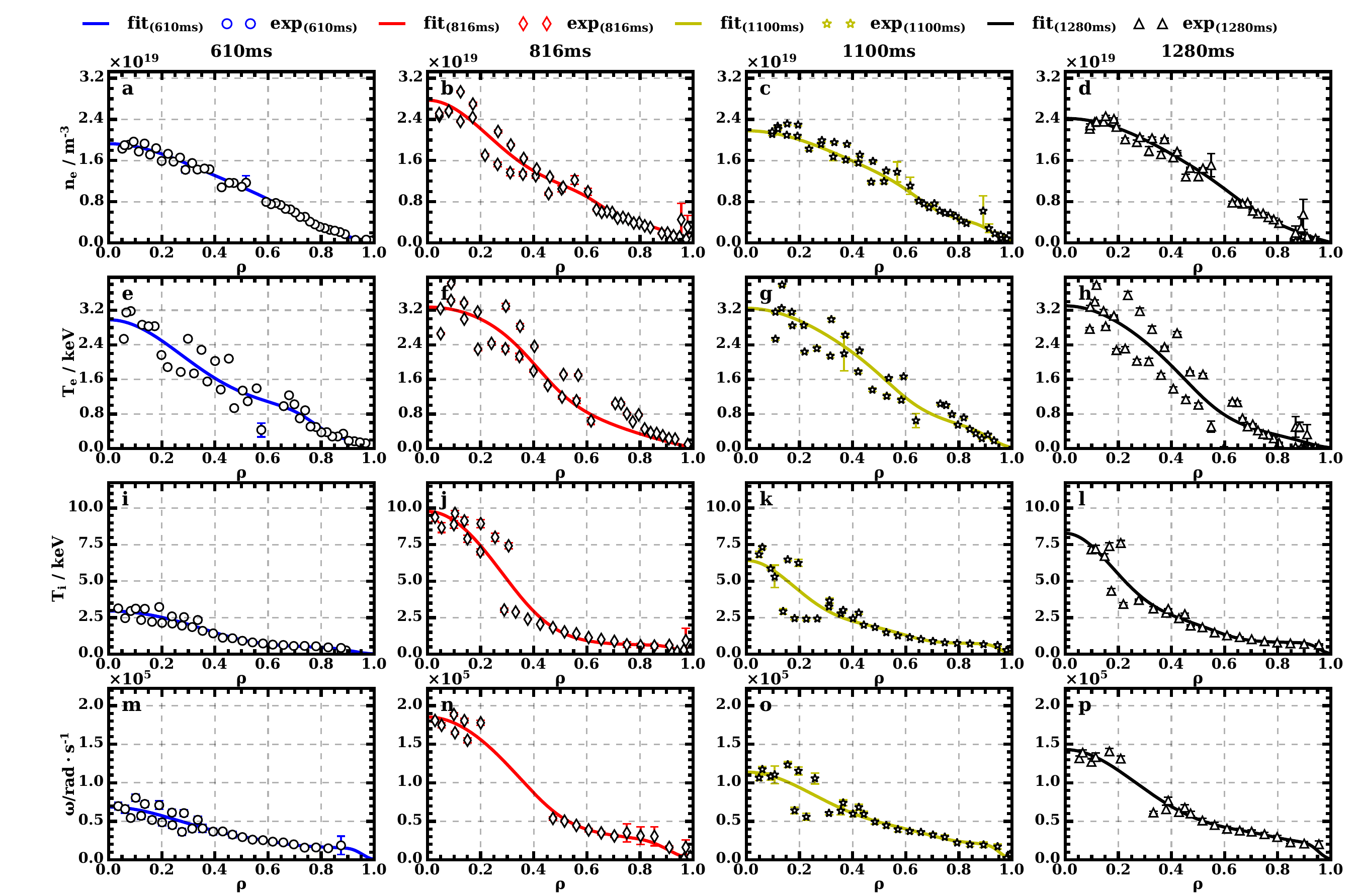}
 \caption{Fitted kinetic profiles at the four time slices, reconstructed from multi-diagnostic synthesis and used as initial inputs to the OMFIT integrated workflow. The rows correspond to electron density $n_e$, electron temperature $T_e$, ion temperature $T_i$, and ion rotation frequency $\omega$ (top to bottom), and the columns to the four time slices $t_1$, $t_2$, $t_3$, and $t_4$ (left to right). The colored lines are the fits and the markers are the measured points, with error bars giving the experimental uncertainty. The fits stay inside them. The color and marker identify the time slice: $t_1$ (blue, circles), $t_2$ (red, diamonds), $t_3$ (yellow, asterisks), and $t_4$ (black, upward triangles). The radial coordinate is $\rho = r/a$.}
\label{fig5}
\end{figure}

The detailed physical workflow and setup of each modeling module follow the description in Ref.~\cite{shu2026integrated}.
For each of the four time slices, we jointly fitted the instantaneous experimental data from multiple diagnostics to construct time-resolved kinetic profiles. This avoids using profiles from a single steady state snapshot. The four slices mark the fishbone growth phase ($t_1$), the nonlinear saturation ($t_2$), the sawtooth ITB decay ($t_3$), and the sawtooth ITB recovery ($t_4$). The sawtooth crash develops faster than the time resolution of the CXRS and Thomson scattering (TS) diagnostics. The profiles reconstructed at $t_3$ and $t_4$ carry larger uncertainty than those at $t_1$ and $t_2$. The fishbone-driven ITB collapse in this period is additionally constrained by the experimental measurements (SXR tomography and Mirnov spectrograms) in Section~2. The loop advances equilibrium, sources, and transport in turn: EFIT~\cite{lao1985separation} supplies the magnetic geometry, ONETWO~\cite{stjohn1994transport} the heating and current-drive sources, and TGYRO~\cite{pan2017investigation} the radial fluxes (see Fig.~\ref{fig20} in the Appendix). TGYRO obtains its turbulent fluxes from the quasilinear Trapped Gyro-Landau-Fluid (TGLF) model~\cite{waltz1997gyro}, closed with the SAT2 saturation rule~\cite{staebler2005gyro} on a 21-point radial mesh spanning $0.1 < \rho < 0.9$, and its neoclassical fluxes from NEO~\cite{marin2025full,dudkovskaia2023drift}. In this work TGYRO runs in a flux-only mode: it calls TGLF and NEO to evaluate the turbulent and neoclassical transport coefficients and flux distributions, and it does not iterate the kinetic profiles toward its own steady state; that profile advance is carried by the outer loop. After convergence, TGLF-scan~\cite{waltz1997gyro} maps the linear growth rate against normalized poloidal wavenumber and sweeps selected profile quantities one parameter at a time. Table~\ref{tab1} lists the discharge parameters and compares the OMFIT-simulated plasma parameters against the experimental measurements, including the confinement factor $H_{98}$~\cite{iter1999physics} and the Greenwald density fraction $f_{\mathrm{G}}$~\cite{greenwald2002density}. Figure~\ref{fig5} shows the fitted kinetic profiles at the four time slices.

\begin{table}[htbp]
\caption{Discharge parameters and OMFIT-simulated plasma parameters for HL-3 shot \#11972 at the four time slices. Each entry gives the experimental and simulated values as Exp./Sim.}
\label{tab1}
\begin{center}
\begin{tabular}{c c c c c}
\toprule
Physical parameter & $t_1$(Exp./Sim.) & $t_2$(Exp./Sim.) & $t_3$(Exp./Sim.) & $t_4$(Exp./Sim.) \\
\midrule
$R$ (m) & 1.78/1.78 & 1.78/1.78 & 1.78/1.78 & 1.78/1.78 \\
$a$ (m) & 0.589/0.589 & 0.586/0.586 & 0.589/0.589 & 0.585/0.585 \\
$I_{\mathrm{p}}$ (MA) & 0.609/0.609 & 0.616/0.616 & 0.621/0.621 & 0.621/0.621 \\
$B_{\mathrm{T}}$ (T) & 1.67/1.67 & 1.67/1.67 & 1.67/1.67 & 1.66/1.66 \\
$P_{\mathrm{NBI}}$ (MW) & 1.9/1.9 & 2.5/2.5 & 2.5/2.5 & 2.5/2.5 \\
$\bar{n}_{\mathrm{e}}$ ($10^{19}\,\mathrm{m}^{-3}$) & 1.07/1.11 & 1.37/1.37 & 1.32/1.32 & 1.40/1.41 \\
$W_{\mathrm{E}}$ (MJ) & 0.128/0.130 & 0.184/0.205 & 0.184/0.202 & 0.191/0.188 \\
$\beta_{\mathrm{N}}$ & 0.76/0.84 & 1.17/1.27 & 1.21/1.29 & 1.27/1.24 \\
$H_{98}$ & --/0.62 & --/0.99 & --/0.85 & --/0.94 \\
$q_{\min}$ & --/1.08 & --/1.04 & --/0.96 & --/0.94 \\
$q_{95}$ & --/4.82 & --/5.33 & --/4.76 & --/4.51 \\
$f_{\mathrm{bs}}$ (\%) & --/8 & --/17 & --/13 & --/14 \\
$f_{\mathrm{G}}$ (\%) & --/20 & --/24 & --/23 & --/24 \\
\bottomrule
\end{tabular}
\end{center}
\vspace{1pt}
\end{table}

\begin{figure}[htbp]
 \centering
        \includegraphics[width=0.95\textwidth]{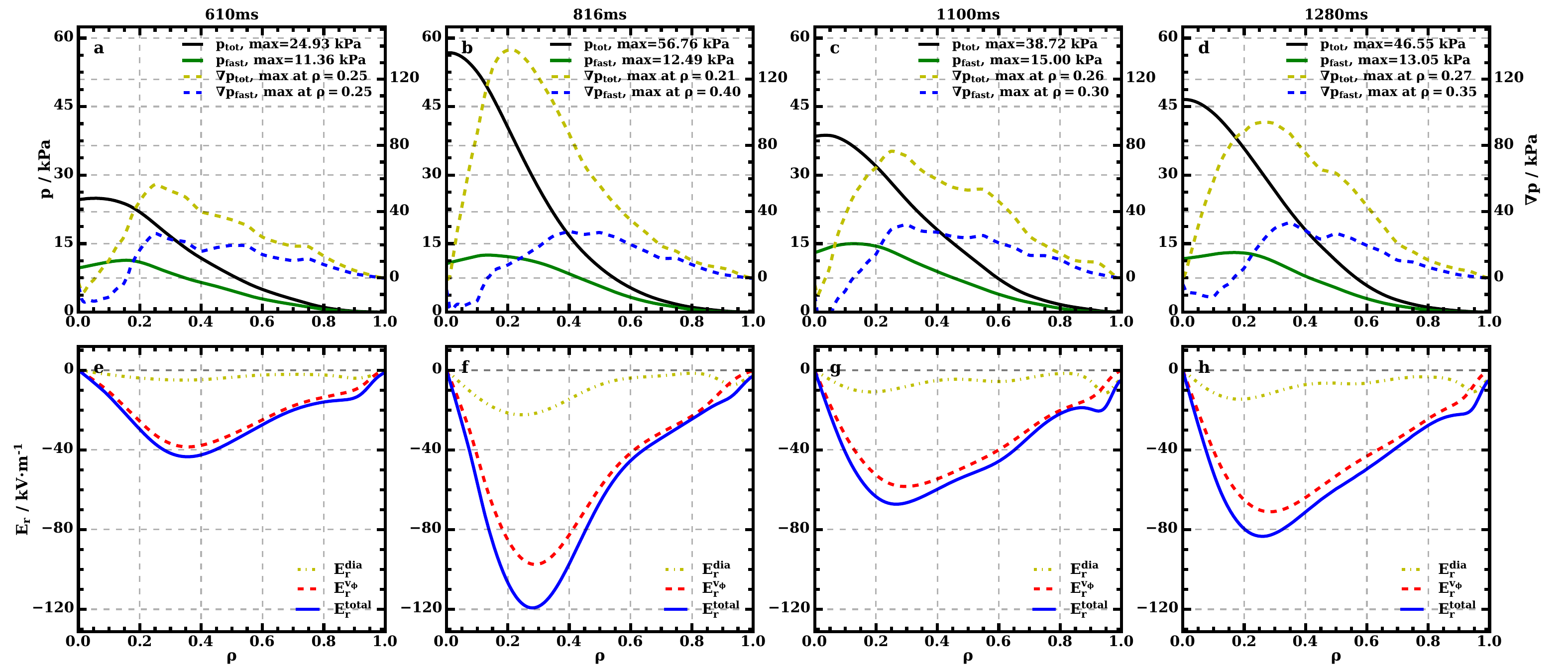}
 \caption{Pressure and radial electric field profiles at the four time slices from the OMFIT workflow. (a)--(d) Total pressure $p_{\mathrm{tot}}$ (black) and fast ion pressure $p_{\mathrm{fast}}$ (green) on the left axis, with their gradients $\nabla p_{\mathrm{tot}}$ (yellow) and $\nabla p_{\mathrm{fast}}$ (blue) on the right axis. (e)--(h) Split of the radial electric field into diamagnetic $E_r^{\mathrm{dia}}$ (yellow), toroidal rotation $E_r^{v_\phi}$ (red), and total $E_r^{\mathrm{total}}$ (blue) parts.}
\label{fig6}
\end{figure}

The radial electric field $E_r$ is obtained from the radial force balance, which splits into diamagnetic, toroidal-rotation, and poloidal-rotation contributions~\cite{shu2026integrated, burrell1999tests}. Figure~\ref{fig6} shows that the toroidal rotation term $E_r^{v_\phi}$ dominates at every time slice, exceeding the diamagnetic term by up to a factor of five. The \(E\times B\) shearing rate follows from the Hahm--Burrell result~\cite{hahm1995flow}, specialized to the large-aspect-ratio tokamak~\cite{hahm1994rotation}
\begin{equation}
\gamma_{E\times B} = \frac{r}{q}\frac{\partial}{\partial r}\left(\frac{q}{r}\frac{E_r}{B}\right),
\label{eq:gamma_EB}
\end{equation}
which, once the toroidal rotation term prevails, gives \(\gamma_{E\times B} \approx (r/q)\,d\omega_{\phi}/dr\). The rotation profile and its radial gradient therefore determine the shearing rate, and whenever it exceeds the drift-wave growth rate the turbulence is quenched, contributing to the ITG stabilization analyzed in Section~5.

Panels (e)--(h) of figure~\ref{fig6} break the radial electric field into the terms that together build the $E\times B$ shear layer.
The diamagnetic piece \(E_r^{\mathrm{dia}}\) never exceeds $20$~kV/m in magnitude. The toroidal-rotation piece \(E_r^{v_\phi}\) carries essentially all of the field. At \(t_2\) (fishbone nonlinear saturation, strongest ITB) it is five times the diamagnetic piece. The peak (most negative) values rank \(t_2\) ($-95$~kV/m) $>$ \(t_4\) ($-70$~kV/m) $>$ \(t_3\) ($-55$~kV/m) $>$ \(t_1\) ($-40$~kV/m), following the rotation history in figure~\ref{fig4}b. Rotation supplies most of the radial electric field and, through it, the $E\times B$ shear that suppresses turbulence.
The peak of \(E_r^{\mathrm{total}}\) lies at $\rho = 0.3$--$0.4$ at both \(t_2\) and \(t_3\), coinciding with the iITB (Fig.~\ref{fig4}a). Inside the ITB the elevated temperature ratio is the main stabilizing contribution, and the rotation-driven $E\times B$ shear grows to comparable importance at the barrier edge, where the temperature ratio falls.

\begin{figure}[htbp]
 \centering
        \includegraphics[width=0.95\textwidth]{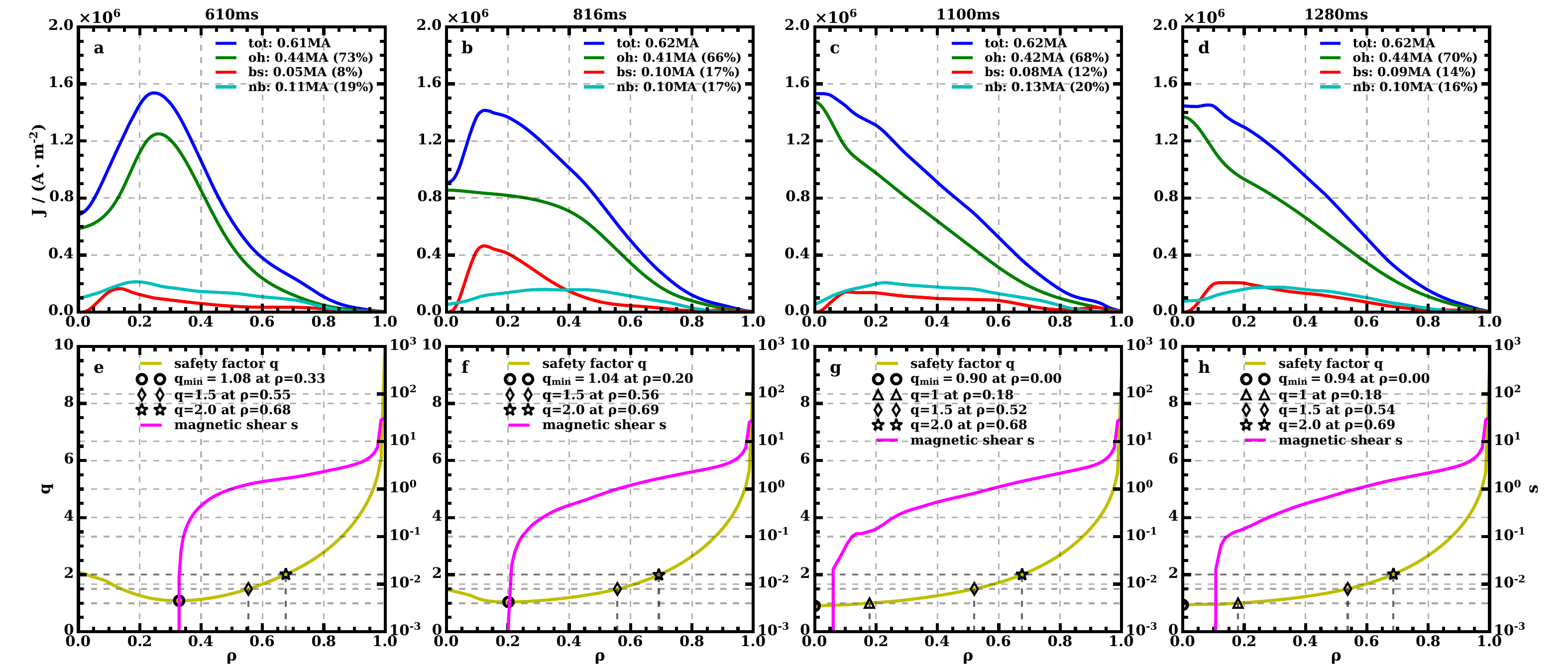}
 \caption{Current, safety factor, and shear profiles at the four time slices from the OMFIT workflow. (a)--(d) Current density: total (blue), Ohmic (green), bootstrap (red), and beam-driven (cyan). (e)--(h) Safety factor $q$ (yellow) and magnetic shear $s$ (magenta). The dashed lines mark $q = 1$, $1.5$, and $2$, with symbols marking the radial location of each crossing of $q_{\min}$, $q = 1$, $q = 1.5$, and $q = 2$.}
\label{fig7}
\end{figure}

The sharply peaked fast-ion pressure profile built by the core NBI deposition (peak $\beta_f \approx 1.14\%$ at $t_3$, Fig.~\ref{fig8}d) supplies the free energy that drives the $m=1$, $n=1$ fishbone instability~\cite{chen1984excitation}. The mode growth and nonlinear saturation at $t_2$ set up a cross-scale coupling loop in which the fishbone drives fast-ion transport that relaxes $\nabla p_{\mathrm{fast}}$.
Figure~\ref{fig7} shows the evolution of the current-density and radial deposition profiles. The HL-3 ITB develops on a core with weak (near-zero) magnetic shear, and low-density operation permits it to form as the two NBI beams step up to 1.2~MW and then 1.3~MW during the main heating phase. At every time slice the current density peaks away from the axis, within \(0.20 < \rho < 0.33\).
The Ohmic component provides most of the current: 0.44\,MA, 0.39\,MA, 0.40\,MA, and 0.43\,MA, that is 73\%, 63\%, 65\%, and 70\% of the total.
As NBI continues, the bootstrap contribution climbs (0.05\,MA, 0.11\,MA, 0.08\,MA, 0.09\,MA), reaching 8\%, 17\%, 13\%, and 14\% in figure~\ref{fig7}(a)--(d).
The beam-driven current falls off slowly.

Figure~\ref{fig7}(e)--(h) plots the simulated safety factor \(q\) and magnetic shear \(s\) profiles at the four time slices.
The minimum of \(q\) is \(q_{\text{min}} = 1.08\) at \(t_1 = 610\)~ms, \(1.04\) at \(t_2 = 816\)~ms, \(0.96\) at \(t_3 = 1100\)~ms, and \(0.94\) at \(t_4 = 1280\)~ms.
The \(s\) profile follows the same trend, remaining weakly sheared over \(\rho < 0.33\) at \(t_1 = 610\)~ms and \(\rho < 0.20\) at \(t_2 = 816\)~ms (Figs.~\ref{fig7}e--f).
Both match the sawtooth oscillations~\cite{vongoeler1974studies} recorded in the soft X-ray channel (Fig.~\ref{fig1}d) and the current deposition profiles of figure~\ref{fig7}(a)--(d). The high core ion temperature and its long-lived ITB are maintained jointly by low density at high current, weak shear, \(E\times B\) shearing rate $\gamma_{E\times B}$ and strong NBI momentum input.

Across the four time slices, the radial location of $q_{\min}$ moves steadily inward, from $\rho = 0.33$ at $t_1 = 610$\,ms (early ITB formation) to $\rho = 0.20$ at $t_2 = 816$\,ms (fishbone nonlinear saturation). It then drops to $\rho = 0.0$ at $t_3 = 1100$\,ms (post-sawtooth crash) and stays there at $t_4 = 1280$\,ms (ITB recovery). The current density profile changes from an off-axis distribution during early ITB formation to an on-axis peaked distribution. During its active phase the fishbone mode holds $q_{\min}$ slightly above unity through its $m=1,n=1$ internal kink resonance ($q_{\min}=1.08$ at $t_1$, $1.04$ at $t_2$). This $q_{\min}\approx 1$ condition sustains the off-axis current deposition profile and the weak (near-zero) magnetic shear ($s\approx 0$). It reduces the ITG drive and gives a favorable magnetic topology for iITB formation. The fishbone mode helps ITB establishment (holding $q_{\min}\approx 1$ and weak shear), while the sawtooth crash that accompanies it drives the fast-ion redistribution analyzed in Section~4. As $q_{\min}$ drops below unity, the sawtooth crash occurs. During the cyclic process of sawtooth crashes and ITB rebuilding with fishbone mode reappearance, $q_{\min}$ oscillates around $q = 1$.

In JT-60U's strong reversed-shear discharges (\(s<0\), \(|s|\gg0\), \(q_{\min}>2\)), the barrier is trapped at the \(q_{\min}\) surface~\cite{fujita1999high,Sakamoto2001}. HL-3 operates near zero shear, which suppresses the ITG growth rate only mildly. Turbulence suppression must come from the \(E\times B\) shear, and whether it overlaps the ITB (Fig.~\ref{fig5}) determines the outcome. Larger devices with stronger heating shift the ITB-formation physics away from shear reversal and toward rotation-driven flow shear.

\section{NBI deposition, beta, and transport simulation}
\begin{figure}[htbp]
 \centering
        \includegraphics[width=0.85\textwidth]{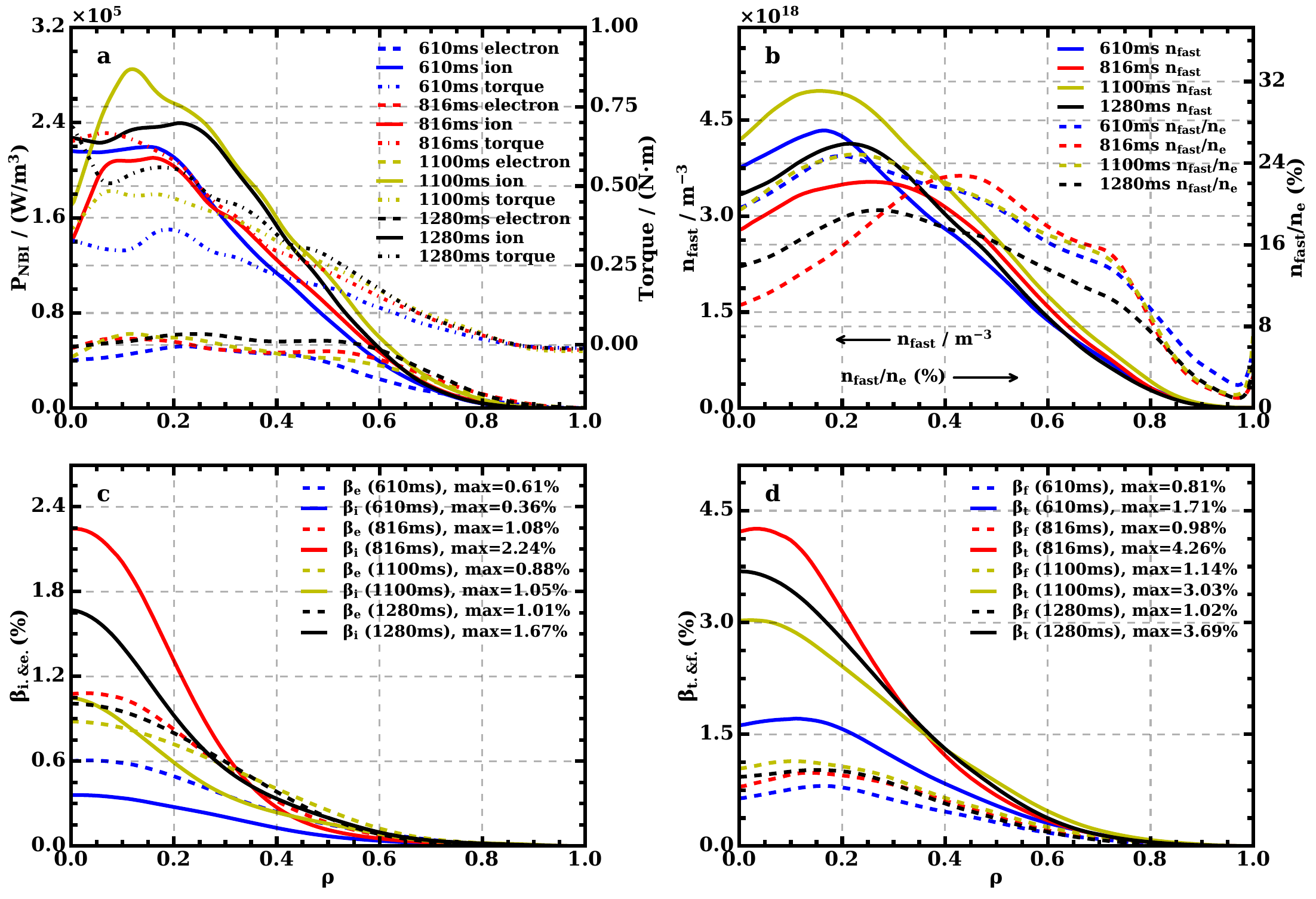}
 \caption{Radial profiles from ONETWO/NUBEAM and TGLF-scan at the four time slices. (a) NBI power density to electrons (dashed) and ions (solid, left axis), and the ion torque density after TGYRO iteration (dash-dotted, right axis). (b) Fast ion density (solid, left axis) and fast-ion-to-electron density ratio (dashed, right axis, \%). (c) Electron beta $\beta_{\mathrm{e}}$ (dashed) and thermal ion beta $\beta_{\mathrm{i}}$ (solid). (d) Fast ion beta $\beta_{\mathrm{f}}$ (dashed) and total beta $\beta_{\mathrm{t}}$ (solid). Blue, red, yellow, and black mark $t_1$, $t_2$, $t_3$, and $t_4$.}
\label{fig8}
\end{figure}

Figure~\ref{fig8}a shows the NBI deposition and fast-ion profiles obtained from ONETWO coupled to \textsc{nubeam}~\cite{pankin2004tokamak}. The NBI heating in this discharge couples mainly and strongly to the ion channel. In the core region ($\rho < 0.4$), the ion heating power density exceeds the electron heating power density by a factor of four to six at all four time slices. This ion-dominated heating follows from the collisional slowing-down of the injected beam ions. A fast ion loses energy through Coulomb collisions with both the electrons and the thermal ions. The share deposited on each species is set by the critical energy $E_c$~\cite{stix1972heating}, the energy at which the electron and ion drag rates are equal. Above $E_c$ the beam heats mainly the electrons, and below $E_c$ it heats mainly the ions. The critical energy is $E_c \approx 14.8\,A_b\,T_e\,(Z_i^2/A_i)^{2/3}$, where $A_b$ is the mass number of the beam ions, $T_e$ is the electron temperature in keV, and $Z_i$ and $A_i$ are the charge and mass numbers of the background thermal ions. For a deuterium beam ($A_b = 2$) injected into a deuterium plasma ($Z_i = 1$, $A_i = 2$) at $T_e \approx 3.3$\,keV, this gives $E_c = 14.8 \times 2 \times 3.3 \times (1/2)^{2/3} \approx 62$\,keV. The NBI beam energy ($45$\,keV) lies below $E_c$. The beam ions transfer most of their energy to thermal ions via collisional drag rather than to electrons. This ion-biased power partition is the main cause of the extreme $T_i/T_e \approx 3$ ratio that marks the HIM. It also ensures that the NBI torque mainly enters the ion rotation channel.

At \(t_1 = 610\)\,ms (early ITB formation), the still-low core density gives a fairly high NBI deposition. Performance peaks at \(t_2 = 816\)\,ms (fishbone nonlinear saturation, strongest ITB).

Figure~\ref{fig8}b shows the fast ion density and its ratio to the electron density. These reflect the cross-scale coupling between fast ion transport and the fishbone mode.
At \(t_1 = 610\)\,ms (early ITB formation) the core density stays low, so the NBI fast ions deposit readily and the fast ion-to-electron density ratio $n_{\mathrm{fast}} / n_{\mathrm{e}}$ is centrally peaked. By \(t_2 = 816\)\,ms (fishbone nonlinear saturation) the mode drives resonant fast ions outward through wave-particle interaction, moving trapped fast ions from the core to \(\rho > 0.4\). Because these ions carry the NBI torque, their redistribution shifts the toroidal momentum to larger radii, decouples the torque source from the core plasma, and pushes the $n_{\mathrm{fast}} / n_{\mathrm{e}}$ peak outward while lowering its on-axis value. At \(t_3 = 1100\)\,ms (post-sawtooth crash) the outer-region fast-ion population stays close to its $t_2$ level and the core value recovers, improving collisional energy transfer to the thermal ions and raising the NBI heating efficiency to its maximum. The sawtooth crash that accompanies the fishbone expels the resonant fast ions, temporarily decouples the torque source from the core, and dominates the drop in toroidal rotation from above $100\,\mathrm{krad/s}$ to $60\,\mathrm{krad/s}$ at $1100$\,ms.
Figure~\ref{fig8}(c),(d) shows the TGLF-scan decomposition of the core beta into its species contributions.
The leading component alternates between fast and thermal ions across the four times. Fast-ion beta $\beta_{\mathrm{f}}$ wins at $t_1 = 610$\,ms ($0.81\%$) and $t_3 = 1100$\,ms ($1.14\%$). Thermal-ion beta $\beta_{\mathrm{i}}$ wins at $t_2 = 816$\,ms and $t_4 = 1280$\,ms, reaching $2.24\%$ at $t_2$. At $t_2$ the fast-ion beta ($0.98\%$) is close to the electron beta ($1.08\%$). The total beta $\beta_{\mathrm{t}}$ peaks at $4.26\%$ at $t_2$, drops to $3.03\%$ at $t_3 = 1100$\,ms as the post-sawtooth ion temperature relaxes and $\beta_{\mathrm{i}}$ falls, then recovers to $3.69\%$ at $t_4 = 1280$\,ms.

\begin{figure}[htbp]
 \centering
        \includegraphics[width=0.8\textwidth]{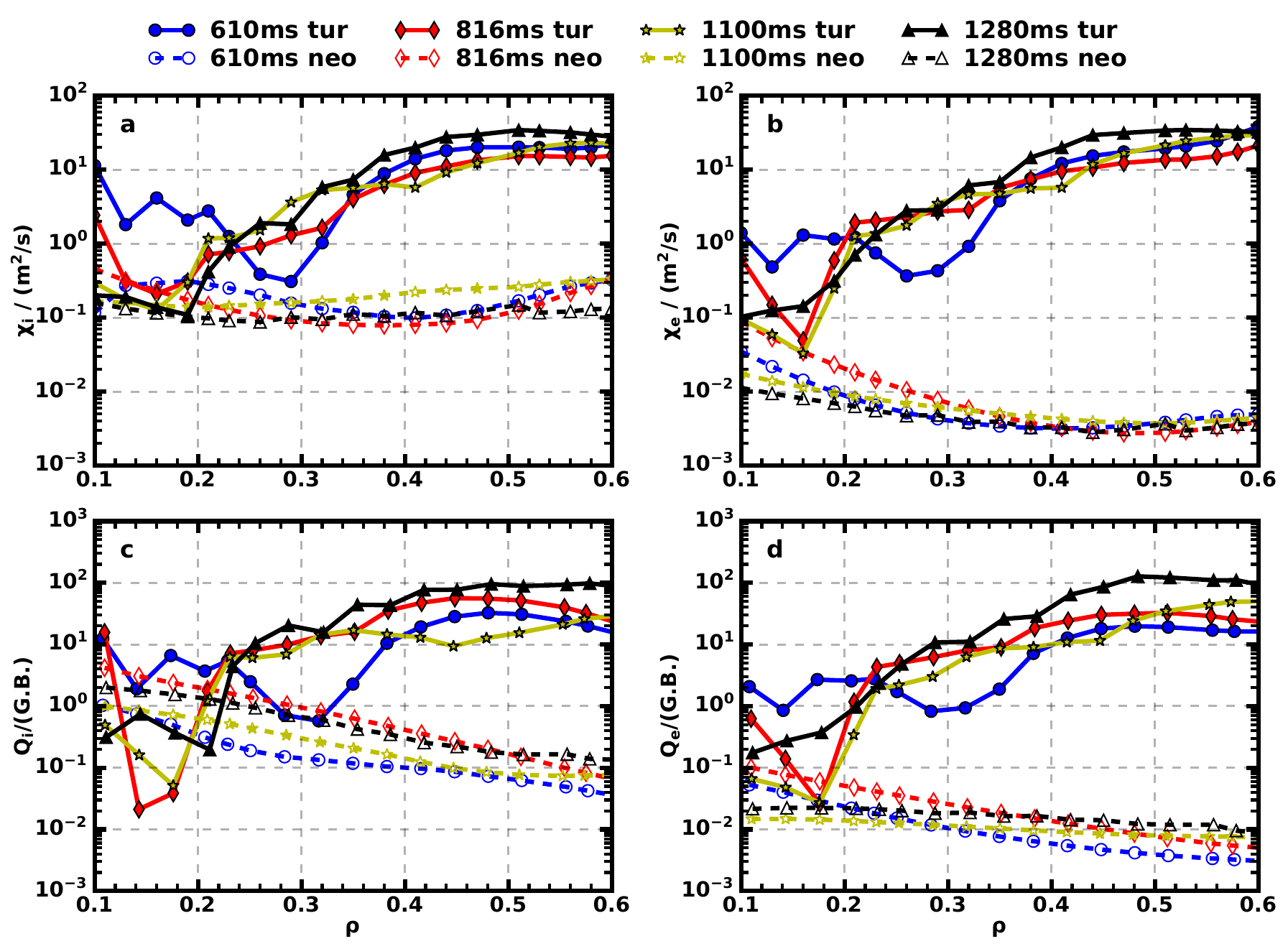}
 \caption{Turbulent and neoclassical transport at the four time slices. (a),(b) Ion and electron thermal diffusion coefficients from TGLF (turbulent) and NEO (neoclassical). (c),(d) Corresponding Gyro-Bohm energy fluxes. Filled markers are turbulent (tur) and open markers neoclassical (neo). Blue, red, yellow, and black mark the four time slices $t_1$--$t_4$.}
\label{fig9}
\end{figure}

Figure~\ref{fig9} shows the turbulent and neoclassical coefficients and Gyro-Bohm-normalized fluxes from TGYRO. Inside $\rho < 0.3$ the fast-ion suppression, the iITB, and the strong $E\times B$ shear jointly push the ion turbulent transport down to the neoclassical level at $t_2$--$t_4$. At $t_1$ it remains far above the neoclassical level. Electron turbulent transport stays about ten times above neoclassical, and the electron scale channel is not fully quenched by the $E\times B$ shear.
In the outer region beyond the core ITB ($\rho > 0.4$), turbulent transport still dominates. The electron and ion turbulent transport coefficients and Gyro-Bohm-normalized energy fluxes there are of comparable magnitude across all four time slices, with the normalized ion heat flux ($Q_i^{\mathrm{turb}}/Q_i^{\mathrm{GB}}$) staying stiffness-locked within $10$--$100$. The ion neoclassical transport coefficient and energy flux exceed the corresponding electron values by more than an order of magnitude. The neoclassical transport is carried almost entirely by the ions. The total transport in this outer region is carried jointly by the ion and electron turbulent channels.

The ion heating power density (Fig.~\ref{fig8}a) rises from $t_1$ to $t_2$ as the total NBI power grows from $1.9$ to $2.5$\,MW at fixed beam energy, and by a further 30\% from $t_2$ to $t_3$. The fishbone-driven outward migration of the fast ions empties the core phase space, and the continuously injected high-energy beam ions deposit their power more readily in the core. The fast ion density (Fig.~\ref{fig8}b) reflects the fishbone-driven transport. At $t_2$ the core ($\rho < 0.25$) drops below its $t_1$ value and the outer region rises, and at $t_3$ continued injection raises the density across $\rho < 0.6$ by a factor of $1.2$. The density peak shifts from $\rho = 0.15$ at $t_1$ and $t_3$ to $\rho = 0.25$ at $t_2$, when the core ion temperature and ITB strength peak, marking the outward redistribution that relaxes the core fast-ion pressure gradient.

The ion torque density (Fig.~\ref{fig8}a, right axis) varies most strongly. Toroidal momentum conservation forces the diffusive momentum flux to vanish on axis, concentrating the effective torque there.
This torque-source relocation, not any loss of total NBI momentum, decouples the NBI torque from the core. The fishbone-driven outward shift of the fast-ion density peak (Fig.~\ref{fig8}b) moves the point where the torque enters the thermal channel, while the large-scale sawtooth crash collapses the core rotation from above $100$ to $60$~krad/s. The NBI torque drives the core rotation and $E\times B$ shear that sustain the ITB. The same fast-ion population supplies the steep $\nabla p_{\mathrm{fast}}$ that excites the $m=1,n=1$ fishbone, and this steep gradient also damps the ion-scale ITG, so the fishbone helps to sustain rather than erode core confinement. MHD excitation, fast-ion redistribution, and microscale transport bifurcation form one cross-scale mechanism.

\section{Micro-turbulence instability analysis}
\begin{figure}[htbp]
 \centering
        \includegraphics[width=0.8\textwidth]{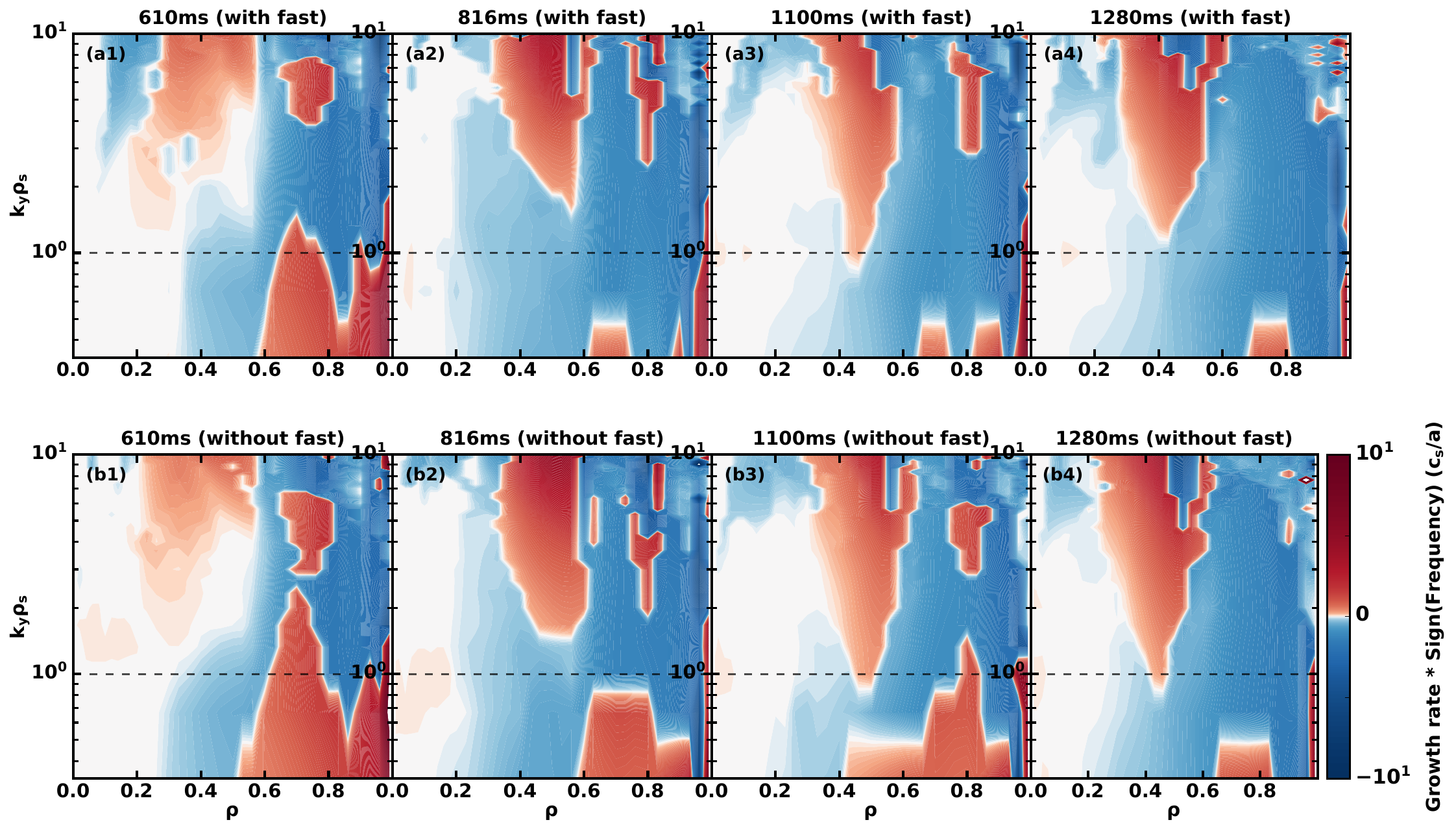}
 \caption{Drift wave turbulence instability spectra of the most unstable eigenmode from TGLF-scan simulations as functions of the normalized poloidal wavenumber $0.3 < k_y \rho_s < 10$ and radial position $0 < \rho < 1$. The columns correspond to the four time slices $t_1$, $t_2$, $t_3$, and $t_4$. The first row (a1)--(a4) shows the cases with fast ions and the second row (b1)--(b4) the cases without fast ions. Blue indicates the ion diamagnetic drift direction, red the electron diamagnetic drift direction, and darker coloring higher instability growth rates.}
\label{fig10}
\end{figure}

The linear spectra span two wavenumber ranges: the ion-scale branch ($k_y\rho_s \lesssim 1$, the ITG and TEM modes) and the electron-scale branch ($k_y\rho_s \gtrsim 1$, the ETG mode). The stabilization mechanisms central to this work (the high temperature ratio, fast-ion dilution, and $E\times B$ shear) act on the ion-scale branch, which is therefore the focus here. The electron-scale ETG mode is identified in the spectra but is not quenched by these ion-scale mechanisms and lies outside the cross-scale coupling of Section~4.
Figure~\ref{fig10} maps, for each of the four time slices, the linear instability spectrum versus normalized poloidal wavenumber at several radii, as computed with TGLF-scan. The spectra show that in the intermediate radial region \(0.3 < \rho < 0.6\), the electron temperature gradient (ETG) mode dominates at all four time slices. It propagates in the electron diamagnetic drift direction. At \(t_1 = 610\)\,ms, during the early nonlinear growth phase of the fishbone, the trapped electron mode (TEM) appears in the outer region \(\rho > 0.6\). At \(t_2 = 816\)\,ms, the fishbone nonlinear saturation phase, the TEM weakens. The instability turns to the ion temperature gradient (ITG) mode in the ion diamagnetic drift direction. At \(t_3 = 1100\)\,ms (post-sawtooth crash), the redistribution of core fast ions improves the core microturbulence stability.

The dominance of the ion direction ITG over the electron direction TEM in $\rho > 0.6$ follows from the asymmetry of the drive sources there. This region lies outside the ion temperature ITB, where the ion temperature gradient drops sharply (the $T_i$ row of Fig.~\ref{fig5}) and weakens the direct ITG drive. The electron density and temperature gradients remain substantial (the $n_e$ and $T_e$ rows of Fig.~\ref{fig5}) and keep driving the electron direction TEM. Without the fast ions, the TEM dominates. Once the fast ions are added, their gradual outward diffusion at $t_2$ and $t_3$ (Fig.~\ref{fig3}) stabilizes the TEM to a degree that varies with the local fast-ion population. The TEM is stabilized and the ion direction ITG becomes the dominant unstable branch. The fast ions stabilize the TEM preferentially. The TEM is driven by trapped electrons~\cite{horton1999drift}, and the fast ions occupy the same trapped phase space. Their precessional dynamics couple directly to the trapped-electron resonance and damp the electron direction mode~\cite{na2025fast, di2021new}. The ITG, driven by the thermal-ion temperature gradient, is affected only through the weaker dilution channel.

\begin{figure}[htbp]
 \centering
        \includegraphics[width=0.8\textwidth]{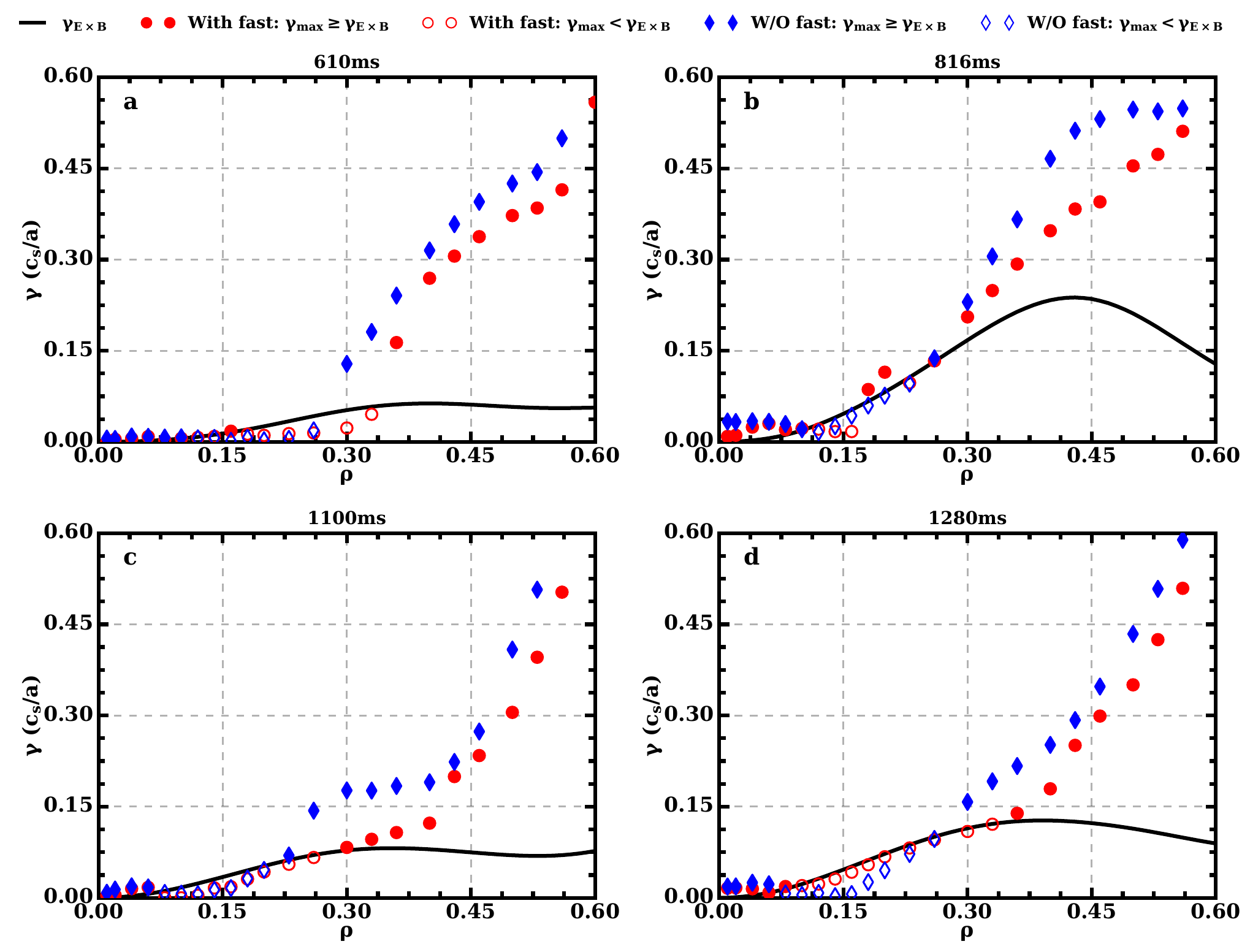}
 \caption{TGLF-scan comparison of the maximum linear growth rate $\gamma_{\max}$ and the $E\times B$ flow shearing rate $\gamma_{E\times B}$ in the core region ($\rho<0.6$) at four time slices: (a) $t_1$, (b) $t_2$, (c) $t_3$, and (d) $t_4$. In each panel, diamonds denote the case without fast ions (blue) and circles the case with fast ions (red). Filled symbols mark $\gamma_{E\times B} < \gamma_{\max}$ and open symbols mark $\gamma_{E\times B} > \gamma_{\max}$.}
\label{fig11}
\end{figure}

To quantify the fast ion effect on core microturbulence, we compared the TGLF-scan maximum linear growth rate $\gamma_{\max}$ and the $E\times B$ shearing rate $\gamma_{E\times B}$ in the core ($\rho<0.6$) at the four time slices, with and without fast ions (Fig.~\ref{fig11}). The fast ion population consistently lowers $\gamma_{\max}$ across all four slices, confirming that fast ions stabilize ion scale drift-wave turbulence rather than acting as an inert background, in agreement with the improved core stability at $t_3$ after fast ion redistribution. Part of this stabilization is dilution: the fast ions replace thermal ions in the ITG drive, weakening the instability for a given ion pressure gradient. This dilution channel complements the temperature-ratio and $E\times B$ shear stabilization discussed below.
SXR tomography (Fig.~\ref{fig3}) shows that the fishbone mode continuously moves resonant fast ions outward through its nonlinear saturation cycle. The TGLF spectra (Fig.~\ref{fig10}) confirm the microturbulence response. At $t_2 = 816$\,ms the fast ions moved to $\rho > 0.4$ leave the TEM weakened and the ion direction ITG appearing in the outer region ($\rho > 0.6$). At $t_3 = 1100$\,ms the post-sawtooth re-accumulation of core fast ions restores the stabilization.

\begin{figure}[htbp]
 \centering
        \includegraphics[width=0.8\textwidth]{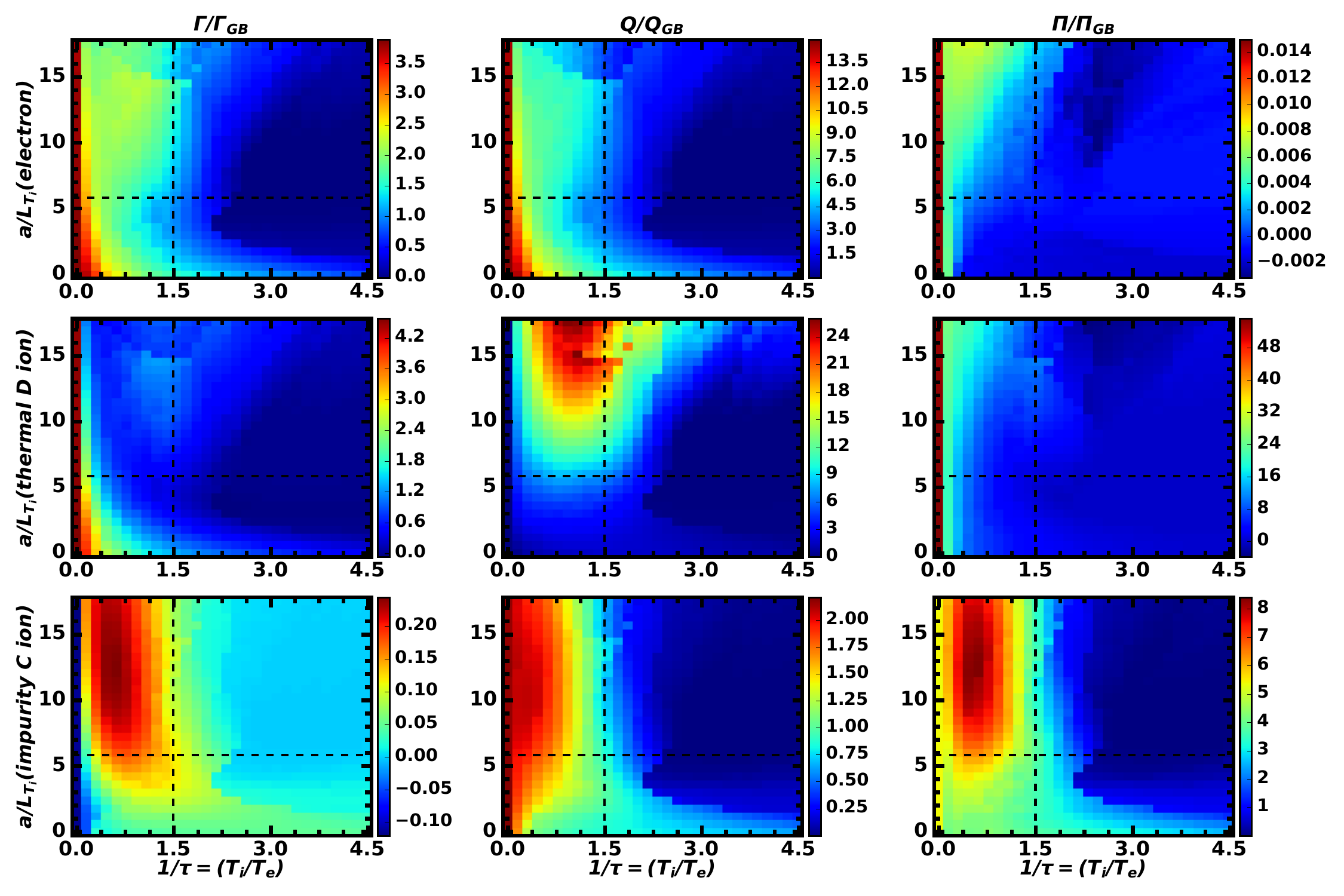}
 \caption{TGLF two-dimensional flux scan of the Gyro-Bohm-normalized particle ($\Gamma/\Gamma_{\mathrm{GB}}$), energy ($Q/Q_{\mathrm{GB}}$), and momentum ($\Pi/\Pi_{\mathrm{GB}}$) fluxes (columns) for the electron, thermal deuterium, and impurity carbon ion species (rows) at $t_2 = 816$\,ms and $\rho = 0.4$. The horizontal axis is the ion-to-electron temperature ratio $1/\tau = T_i/T_e$, and the vertical axis is the normalized ion temperature gradient $a/L_{T_i}$, both scanned over the range $0$ to $3$ times their experimental values. Dashed lines mark the experimental operating point.}
\label{fig12}
\end{figure}

The effect of the hot ion temperature ratio on turbulent transport is isolated with a TGLF two-dimensional scan at $t_2 = 816$\,ms and $\rho = 0.4$, sweeping the normalized ion temperature gradient $a/L_{T_i}$ and the ratio $1/\tau = T_i/T_e$ over $0$ to $3$ times their experimental values (Fig.~\ref{fig12}). The particle, energy, and momentum fluxes of all three species (electron, thermal deuterium, and impurity carbon) fall steadily as $1/\tau$ increases, most steeply between $1$ and $2$ times its experimental value. The high $T_i/T_e$ of the HIM is inherently favorable to high confinement, with no anomalous high-flux band in the scanned range.
In the low-ratio region ($1/\tau$ below its experimental value), Fig.~\ref{fig12} show a direct relation between $1/\tau$ and $a/L_{T_i}$. A larger $a/L_{T_i}$ is needed to reach a given flux as $1/\tau$ rises, for all three transport channels alike. The flux falls as $1/\tau$ rises. This raises the ITG excitation threshold. A larger temperature ratio demands a steeper $a/L_{T_i}$ to drive the same flux. Once the gradient exceeds the raised threshold, the growth rate and flux rise linearly with $a/L_{T_i}$. This mapping between the temperature ratio and the required gradient gives a model-based constraint for future HIM operation.

\begin{figure}[htbp]
 \centering
        \includegraphics[width=0.8\textwidth]{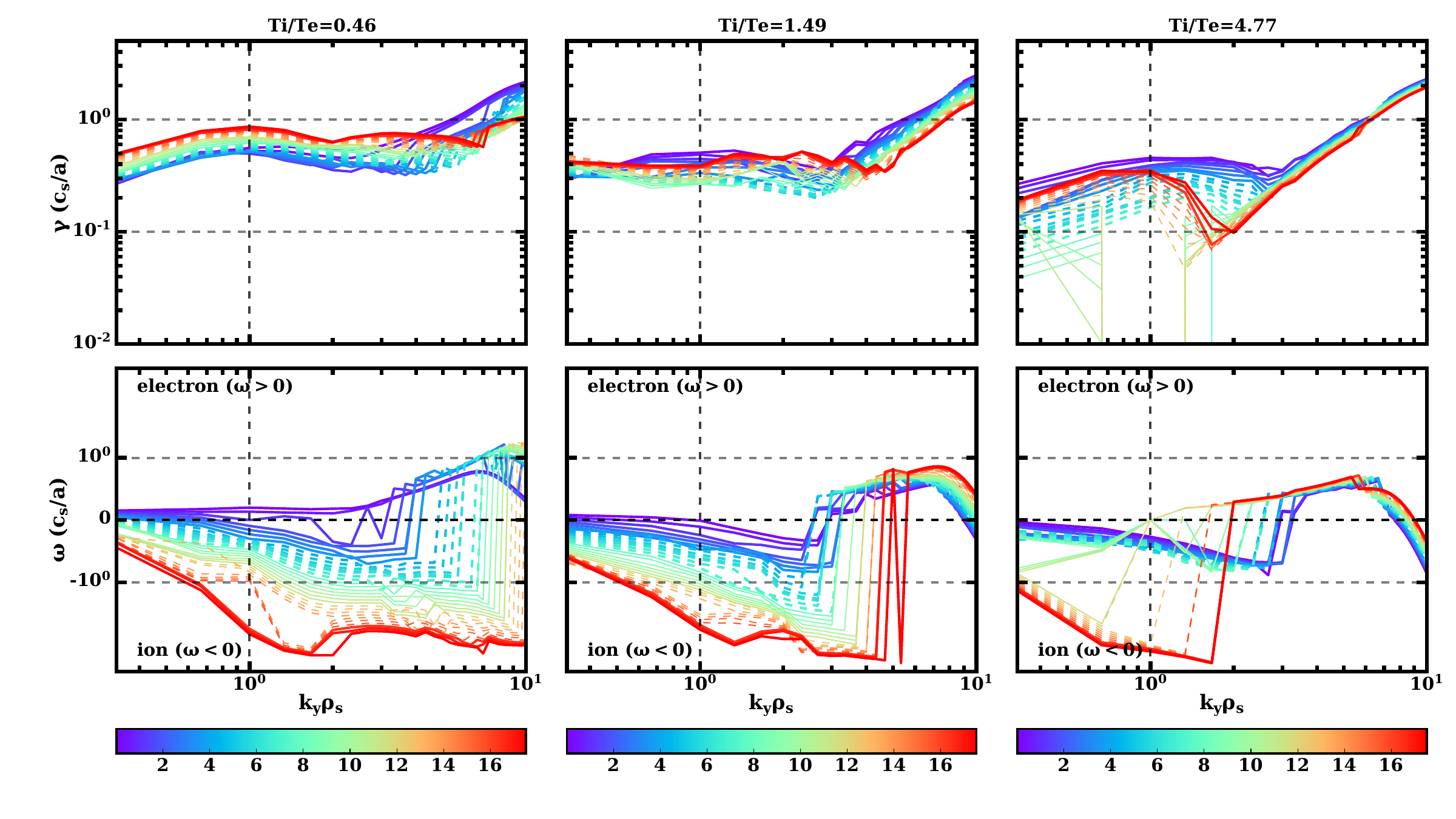}
 \caption{TGLF linear eigenvalue spectra with fast ions at $t_2 = 816$\,ms and $\rho = 0.4$: growth rate $\gamma$ (top row) and real frequency $\omega$ (bottom row), both in units of $c_s/a$, as functions of the normalized poloidal wavenumber $k_y\rho_s$, for three fixed ion-to-electron temperature ratios $T_i/T_e = 0.3$, $1.0$, and $3.0$ times the initial value, corresponding to $T_i/T_e = 0.46$, $1.49$, and $4.77$ (columns). Within each column, the rainbow-colored curves correspond to a one-dimensional scan of the normalized ion temperature gradient $a/L_{T_i}$ over $0$ to $3$ times its initial value, with the horizontal colorbar beneath each column indicating the scanned value. The vertical dashed line marks $k_y\rho_s = 1$, and the horizontal dashed line in the bottom row separates the electron direction ($\omega > 0$) from the ion direction ($\omega < 0$). As $T_i/T_e$ rises from the left to the right column, the high-wavenumber modes move from the ion to the electron diamagnetic direction.}
\label{fig13}
\end{figure}

At $\rho = 0.4$, the scan also locates the ITG excitation threshold. When both $1/\tau$ and $a/L_{T_i}$ lie below their experimental values, the deuterium particle flux does not rise with $a/L_{T_i}$, indicating that $a/L_{T_i} \approx 5.7$ sits near the threshold. This threshold rises linearly with $1/\tau$. The high temperature ratio moves the operating point toward the strongly-suppressed regime.
A complementary one-dimensional TGLF eigenvalue scan at $t_2 = 816$\,ms and $\rho = 0.4$ (Fig.~\ref{fig13}) holds the ion-to-electron temperature ratio fixed at $0.3$, $1.0$, and $3.0$ times its initial value ($T_i/T_e = 0.46$, $1.49$, and $4.77$, one per column) and sweeps $a/L_{T_i}$ over $0$ to $3$ times its initial value. At the lowest ratio ($T_i/T_e = 0.46$) and low gradient ($a/L_{T_i} < 5$), the low-wavenumber drift waves ($k_y\rho_s < 1$) propagate in the electron diamagnetic direction and are identified as trapped electron modes (TEM). As $a/L_{T_i}$ rises they turn into ion direction ion temperature gradient (ITG) modes, and over the scanned range $k_y\rho_s < 5$ the growth rate rises with $a/L_{T_i}$. At the experimental ratio ($T_i/T_e = 1.49$) the raised ITG threshold breaks this trend. The low-wavenumber growth rate no longer rises monotonically with $a/L_{T_i}$ but first falls and then recovers, and even at $3$ times the initial gradient it stays below its value at $0.3$ times the initial gradient. The growth rate in this column is everywhere below that of the $0.3$ times column. At the highest ratio ($T_i/T_e = 4.77$) every low-wavenumber mode ($k_y\rho_s < 1$) propagates in the ion direction, and ITG alone is unstable. The growth rate again falls and then recovers as $a/L_{T_i}$ rises, and even at $3$ times the initial gradient it stays below its value at $0.3$ times the initial gradient. The ion direction ITG dominates at the highest ratio, where the ITG threshold is largest. This appears at first to contradict the threshold formula.

A raised threshold should make excitation harder, and it suppresses the electron direction TEM more strongly than the ion direction ITG. ITG is harder to excite in absolute terms, but it wins the competition against the TEM and emerges as the sole unstable mode. Across the three columns a second trend appears at higher wavenumber. At the lowest ratio the ion-directed branch extends deep into the short-wavelength range, but as $T_i/T_e$ rises its high-wavenumber tail is progressively suppressed and the unstable modes there move to the electron diamagnetic direction, so the crossover wavenumber falls with $T_i/T_e$, confining the ion-directed instability to lower wavenumbers. We interpret this as ion temperature screening. The elevated $T_i/T_e$ of the hot ion mode screens out the ion response at short wavelength, damping the ion-direction ITG and leaving the electron-direction branch to carry the high-$k$ turbulence. The screening then self-consistently stabilizes the ion-direction ITG. The non-monotonic $a/L_{T_i}$ dependence reflects a balance of the ion diamagnetic drive $\omega_{*T}\propto k_\theta\rho_s\,(a/L_{T_i})$, which grows linearly with $k_\theta\rho_s$, against finite-Larmor-radius (FLR) stabilization, which grows quadratically as $(k_\perp\rho_i)^2$ and weakens the drive through the gyroaveraging factor $J_0^2(k_\perp\rho_i)\approx 1-(k_\perp\rho_i)^2/2$. Ion Landau damping strengthens as the parallel phase velocity approaches the ion thermal velocity $v_{ti}=(2T_i/m_i)^{1/2}$, with the Maxwellian resonant rate~\cite{coppi1967instabilities,kadomtsev1970turbulence}
\begin{equation}
\gamma_{\mathrm{LD}}=-\sqrt{\frac{\pi}{2}}\,\frac{\omega^2}{|k_\parallel|v_{ti}}\,\exp\!\left(-\frac{\omega^2}{k_\parallel^2 v_{ti}^2}\right),
\label{eq:landau}
\end{equation}
Both stabilization terms grow faster than the drive with wavenumber and overtake it at high $k_\perp\rho_i$~\cite{horton1999drift,romanelli1989kinetic}, shifting the most unstable mode to shorter wavelength while it stays in the ion direction.

\begin{figure}[htbp]
 \centering
        \includegraphics[width=0.8\textwidth]{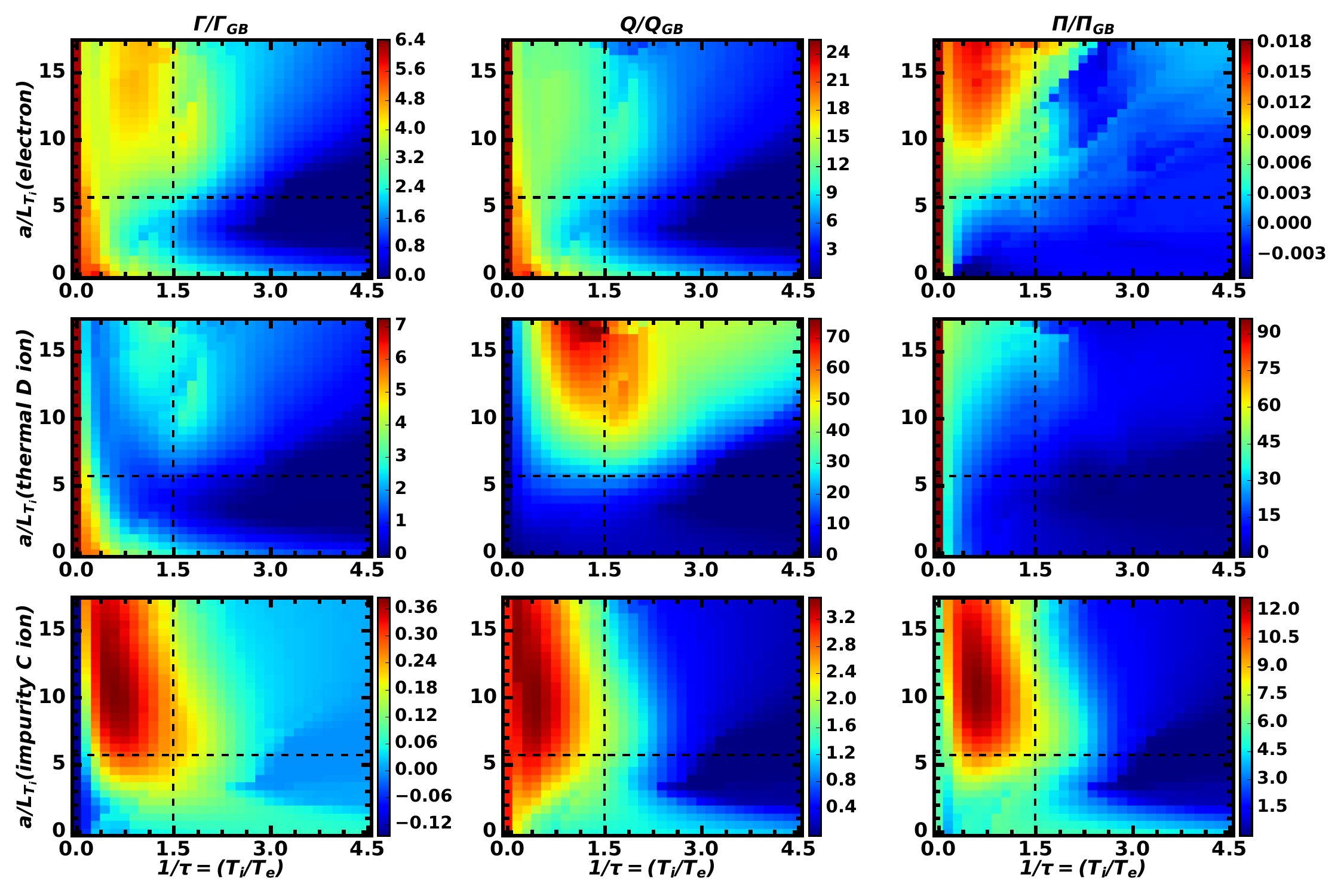}
 \caption{TGLF two-dimensional flux scan identical to Fig.~\ref{fig12} but with the fast-ion population removed.}
\label{fig14}
\end{figure}

The flux decline with $1/\tau$ in Figs.~\ref{fig12} and~\ref{fig14} is a direct consequence of the kinetic ITG threshold. In the flat-density, strong-ballooning limit the toroidal ITG destabilizes above the critical gradient~\cite{horton1999drift,romanelli1989kinetic,hahm1988properties}
\begin{equation}
\left(\frac{R}{L_{T_i}}\right)_c = \frac{4}{3}\left(1+\frac{T_i}{T_e}\right),
\label{eq:itg_threshold}
\end{equation}
a linear increase in $T_i/T_e$. The HIM's large $T_i/T_e$ therefore lifts the critical gradient and curbs ion-scale transport. An explicit fast-ion species raises the threshold further through an extra stabilizing term~\cite{jung2026fast}. The sawtooth crash, synchronized with the fishbone, moves the resonant fast ions outward and temporarily weakens this term, so the threshold falls and the ion direction ITG reappears in the outer region (Fig.~\ref{fig10}), linking the large-scale sawtooth to the microscopic stability threshold.

To separate the fast-ion contribution from the temperature-ratio effect, we repeated the two-dimensional scan with the fast ions removed (Fig.~\ref{fig14}). Relative to Fig.~\ref{fig12}, the fluxes of all three species rise markedly at high $1/\tau$ once the fast ions are absent. Fast ions suppress electron, thermal-ion, and impurity transport alike, most strongly at high $T_i/T_e$. This corroborates the instability-level result of Fig.~\ref{fig11}, where fast ions reduce $\gamma_{\max}$.
The local correspondence between fast ion density and the dominant instability closes the cross-scale coupling of Section~4. Fast ions drive the macroscopic fishbone via $\nabla p_{\mathrm{fast}}$ and suppress microscopic ITG via linear wave-particle interaction. The large-scale sawtooth crash, synchronized with the fishbone, redistributes the fast ions outward and collapses the rotation shear, so the core ITG re-excites and the confinement degrades from $t_2$ to $t_3$ ($T_i(0)$ from $9.8$ to $6.4$\,keV and $H_{98}$ from $0.99$ to $0.85$). The near-neoclassical ion transport at $t_3$ (Fig.~\ref{fig9}) is confined to the deep core ($\rho<0.3$), where fast ions re-accumulate after the crash. The re-excited ITG appears in the outer region ($\rho\approx0.8$) from which the sawtooth-driven redistribution has removed the fast ions.

\begin{figure}[htbp]
 \centering
        \includegraphics[width=0.8\textwidth]{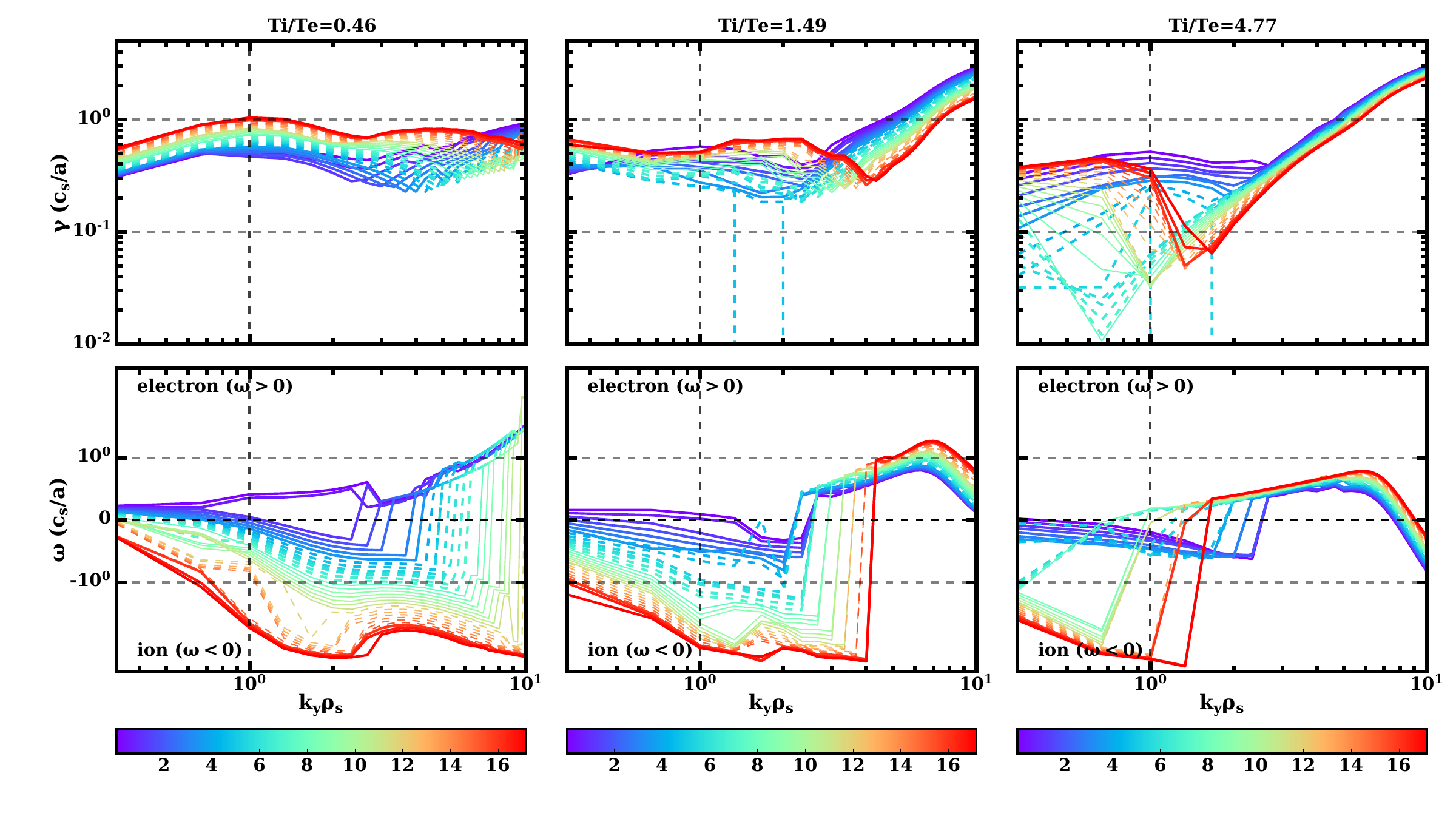}
 \caption{TGLF linear eigenvalue spectra identical to Fig.~\ref{fig13} but with the fast-ion population removed, at $t_2 = 816$\,ms and $\rho = 0.4$.}
\label{fig15}
\end{figure}

Repeating the eigenvalue scan with the fast ions removed (Fig.~\ref{fig15}) leaves the qualitative picture unchanged and raises the growth rate at every temperature ratio. The electron direction TEM, the ion direction ITG, their non-monotonic $a/L_{T_i}$ dependence, and the high-wavenumber ion-to-electron shift all persist without the fast ions, so the ion temperature screening is intrinsic to the temperature ratio rather than the fast-ion population. The overall growth rate is systematically higher than in the fast-ion-inclusive case of Fig.~\ref{fig13}. The suppression is largest at high $T_i/T_e$ and high $a/L_{T_i}$, where the fast-ion-inclusive scan is lower by the widest margin. The elevated temperature ratio and the fast-ion population suppress drift-wave turbulence together, neither alone achieving the depth of suppression they deliver jointly. This cooperative effect underlies the favorable confinement of the HIM and suggests the HIM with its high fast-ion fraction is favorable for high-parameter plasmas, in which the fast-ion population suppresses drift-wave anomalous transport and improves confinement.

\begin{figure}[htbp]
 \centering
        \includegraphics[width=0.8\textwidth]{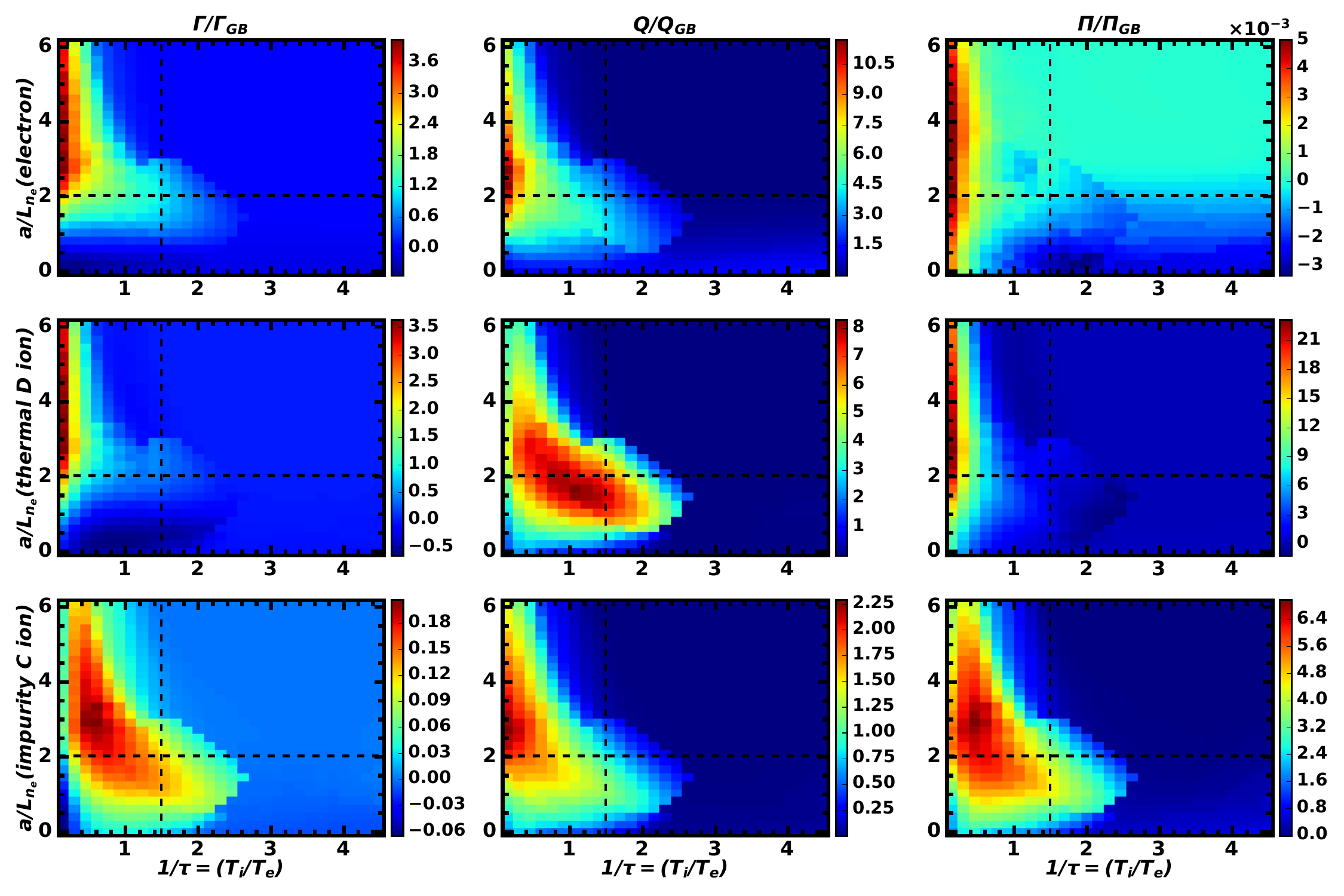}
 \caption{TGLF two-dimensional flux scan identical to Fig.~\ref{fig12} but with the fast-ion population retained and the normalized electron density gradient $a/L_{n_e}$ on the vertical axis in place of $a/L_{T_i}$.}
\label{fig16}
\end{figure}

Figure~\ref{fig16} sweeps the ion-to-electron temperature ratio $T_i/T_e$ against the normalized electron density gradient $a/L_{n_e}$ at $\rho = 0.4$ for $t_2 = 816$\,ms, with the fast-ion population retained. The particle flux of every species falls sharply as $T_i/T_e$ rises, and the suppression is strongest for the main ions: above $T_i/T_e > 1$ the radial deuterium particle transport is essentially quenched. Substantial particle transport appears only for $a/L_{n_e} > 1$, and the impurity carbon flux stays well below the electron and main-ion levels, which are comparable.

The three species share a similar energy-flux threshold. Energy transport is nearly suppressed for $T_i/T_e > 2.6$ and is largest in the $a/L_{n_e} < 3$ band, where the electron and main-ion fluxes are comparable (about $10$ in Gyro-Bohm units) and roughly four times the carbon value (about $2.5$). For $T_i/T_e < 2.6$ the energy flux drops again once $a/L_{n_e}$ exceeds $3$. The momentum flux is carried mainly by the main ions for $T_i/T_e < 1$ and by the impurity ions for $1 < T_i/T_e < 2.6$, a species balance that mirrors the energy flux. Its largest level also sits in the $a/L_{n_e} < 3$ band.

The particle-flux result ties the transport scan back to the instability picture of Fig.~\ref{fig10}. The density gradient $a/L_{n_e}$ drives the trapped electron mode, whose non-adiabatic trapped-electron response carries the anomalous particle flux~\cite{horton1999drift}. The appearance of significant particle transport only for $a/L_{n_e} > 1$ marks the density gradient, not the temperature gradient, as the particle-transport drive, consistent with the deuterium particle flux that stays flat as $a/L_{T_i}$ rises near threshold in Fig.~\ref{fig12}. As $T_i/T_e$ grows, the unstable low-wavenumber branch moves from the electron-direction TEM toward the ion-direction ITG (Fig.~\ref{fig13}), and the ion-direction ITG, with nearly adiabatic electrons, produces little particle flux. The retained fast ions damp the TEM preferentially (Fig.~\ref{fig10}), so the low particle flux at high $T_i/T_e$ in Fig.~\ref{fig16} is the transport-level counterpart of the TEM-to-ITG transition.

\begin{figure}[htbp]
 \centering
        \includegraphics[width=0.8\textwidth]{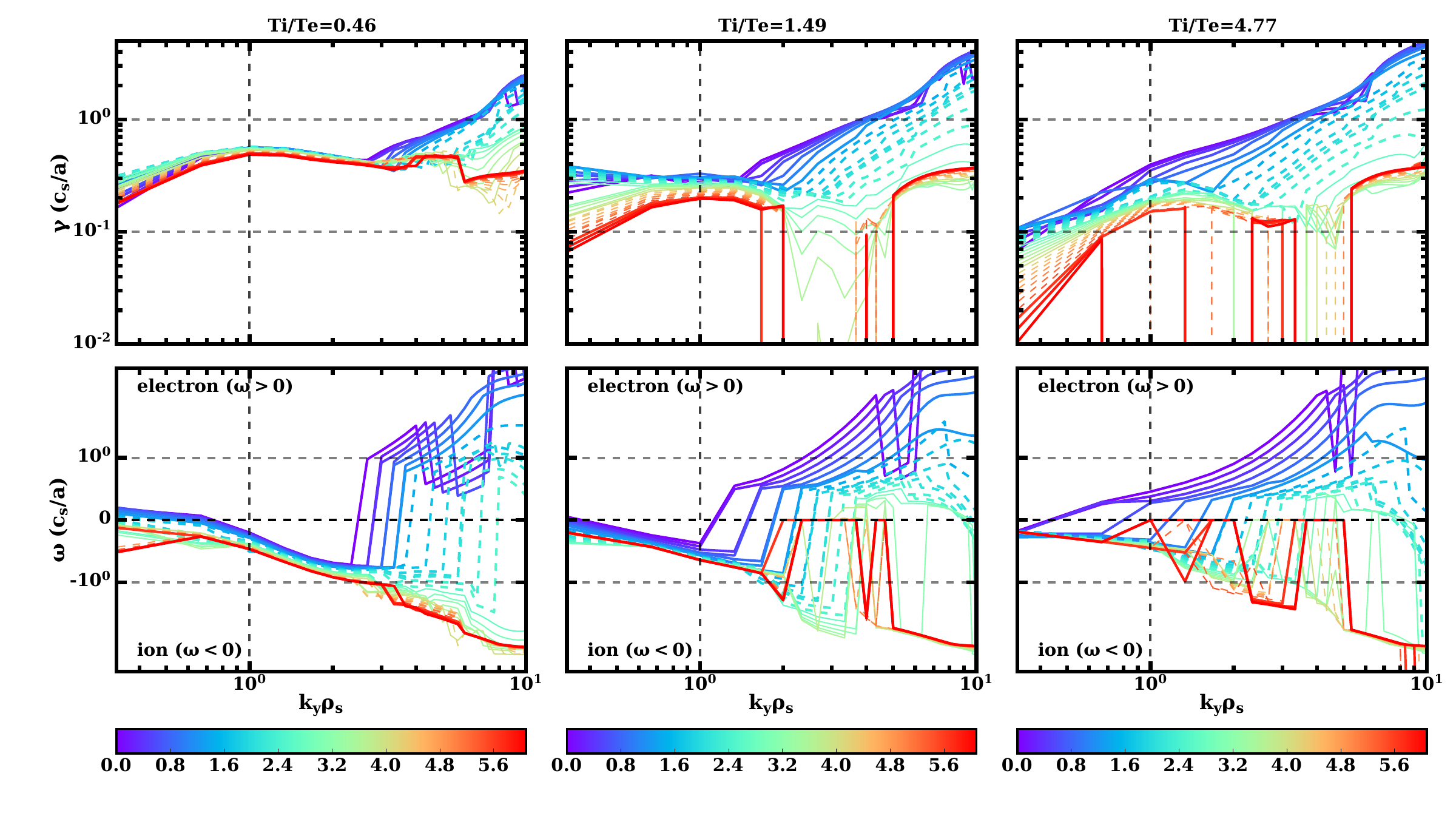}
 \caption{TGLF linear eigenvalue spectra identical to Fig.~\ref{fig13} but with the normalized electron density gradient $a/L_{n_e}$ swept in place of $a/L_{T_i}$. As $a/L_{n_e}$ rises within each column, the high-wavenumber modes move from the electron to the ion diamagnetic direction.}
\label{fig17}
\end{figure}

Figure~\ref{fig17} shows the instability-level counterpart of the density-gradient scan in Fig.~\ref{fig16}. It sweeps the normalized electron density gradient $a/L_{n_e}$ at the three fixed temperature ratios and records the growth rate $\gamma$ and real frequency $\omega$ against $k_y\rho_s$. In the high-wavenumber range ($k_y\rho_s > 1$) the growth rates of both the electron- and the ion-direction branches drop monotonically as $a/L_{n_e}$ rises. The unstable high-wavenumber branch also turns from the electron toward the ion diamagnetic direction, the crossover wavenumber falling so that the ion-directed range widens. A larger $T_i/T_e$ preserves these trends while lowering the growth rate further.

Both trends follow from the way a steeper density gradient dilutes the temperature-gradient drive. The temperature-gradient instabilities are governed by the ratio of the density and temperature scale lengths, the $\eta$ parameter~\cite{coppi1967instabilities,kadomtsev1970turbulence},
\begin{equation}
\eta_s = \frac{L_{n_e}}{L_{T_s}} = \frac{a/L_{T_s}}{a/L_{n_e}}, \qquad s = i, e,
\label{eq:eta}
\end{equation}
with $L_{n_e}$ and $L_{T_s}$ the density and temperature scale lengths. At fixed $a/L_{T_s}$, increasing $a/L_{n_e}$ lowers $\eta_s$ and weakens the temperature-gradient drive, so the growth rate of both branches falls. In the hot ion mode $a/L_{T_i}$ exceeds $a/L_{T_e}$, giving $\eta_i > \eta_e$. As $a/L_{n_e}$ grows, the smaller $\eta_e$ drops below its instability threshold first and quenches the electron-direction branch, while the ion-direction branch, whose larger $\eta_i$ keeps it above threshold, persists to steeper gradients. This differential quenching, not a change in the underlying drive, shifts the high-wavenumber turbulence toward the ion direction in Fig.~\ref{fig17}~\cite{horton1999drift,lee1987collisionless}. A steeper density gradient thus stabilizes the high-wavenumber temperature-gradient branches through $\eta$, apart from its role as the drive of the low-wavenumber trapped-electron particle flux in Fig.~\ref{fig16}.

\begin{figure}[htbp]
 \centering
        \includegraphics[width=0.8\textwidth]{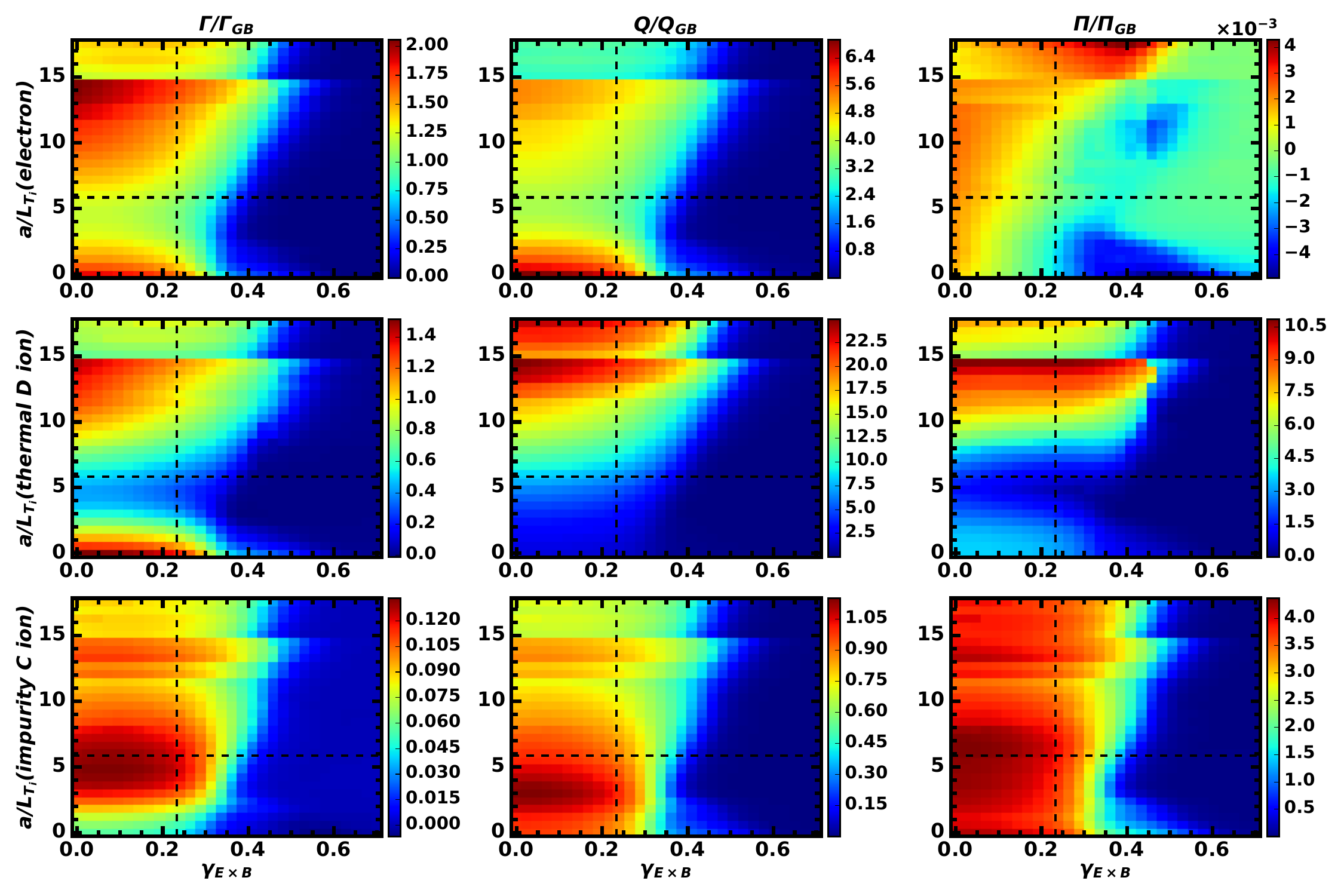}
 \caption{TGLF two-dimensional flux scan identical to Fig.~\ref{fig12} but with the $E\times B$ shear rate $\gamma_{E\times B}$ (in units of $c_s/a$) on the horizontal axis in place of $1/\tau$.}
\label{fig18}
\end{figure}

At the highest temperature ratio ($T_i/T_e = 4.77$, three times the initial value), the ITG growth rate within $k_y\rho_s < 1$ is strongly suppressed inside an optimal window $8 < a/L_{T_i} < 12$, with or without fast ions. Once the iITB is established, a high $T_i/T_e$ combined with a moderate $a/L_{T_i}$ in this range can therefore curb ITG-driven turbulent energy transport and reinforce confinement in future higher-parameter HIM operation. During the early iITB-formation phase the magnetic shear and the $E\times B$ shear are the decisive ingredients, and the fast-ion contribution further amplifies this confinement advantage.
Figure~\ref{fig18} sweeps the $E\times B$ shear rate and the normalized ion temperature gradient $a/L_{T_i}$ at $\rho = 0.4$ for $t_2 = 816$\,ms. As the shear rate rises, the particle, energy, and momentum fluxes are strongly suppressed, and above $\gamma_{E\times B} \approx 0.5\,c_s/a$ (twice the experimental value) they fall to a very low level. The radial transport is effectively controlled.

The three stabilizing channels combine near the threshold, where the maximum linear growth rate rises linearly with the temperature gradient~\cite{romanelli1989kinetic,hahm1988properties}. The eigenvalue scan of Fig.~\ref{fig13} confirms this rise, reaching $\gamma_{\max}\approx0.4\,c_s/a$ at the upper end of the scanned range. The $E\times B$ shear quenches the mode once $\gamma_{E\times B} > \gamma_{\max}$~\cite{hahm1995flow} (the boundary between filled and open symbols in Fig.~\ref{fig11}), so a stronger shear demands a steeper gradient to keep the mode above the quench threshold. The kinetic threshold of Eq.~\eqref{eq:itg_threshold} rises with $T_i/T_e$, fast-ion dilution raises it further~\cite{jung2026fast}, and the shear quench supplies a third, rate-based condition. Raising $\gamma_{E\times B}$ therefore suppresses the fluxes until, above $\gamma_{E\times B}\approx0.5\,c_s/a$, they collapse (Fig.~\ref{fig18}). The same reasoning shows why the ITB is self-sustaining: the steep $\nabla T_i$ is itself an ITG drive, but the NBI-torque-driven shear, the fast-ion dilution, and the elevated temperature ratio together hold the operating point below threshold.

The radial balance of the three channels follows from the measured profiles. Inside the ITB the elevated $T_i/T_e$ dominates the ITG suppression, and the fast-ion dilution strengthens it further. The $E\times B$ shear is smaller in the deep core and becomes the main stabilizing contribution where $T_i/T_e$ drops steeply at the barrier edge, keeping the operating point sub-critical inside the barrier. The sawtooth-affected slices $t_3$ and $t_4$ reverse this: the ITB collapse flattens the core $T_i/T_e$, and the rotation collapse flattens the shear profile, so the ion direction ITG reappears over a broad range. The large-scale sawtooth, not the fishbone, is thus what temporarily removes the stabilization that sustains the ion temperature ITB.

\section{Summary and discussion}

\begin{figure}[htbp]
 \centering
        \includegraphics[width=0.85\textwidth]{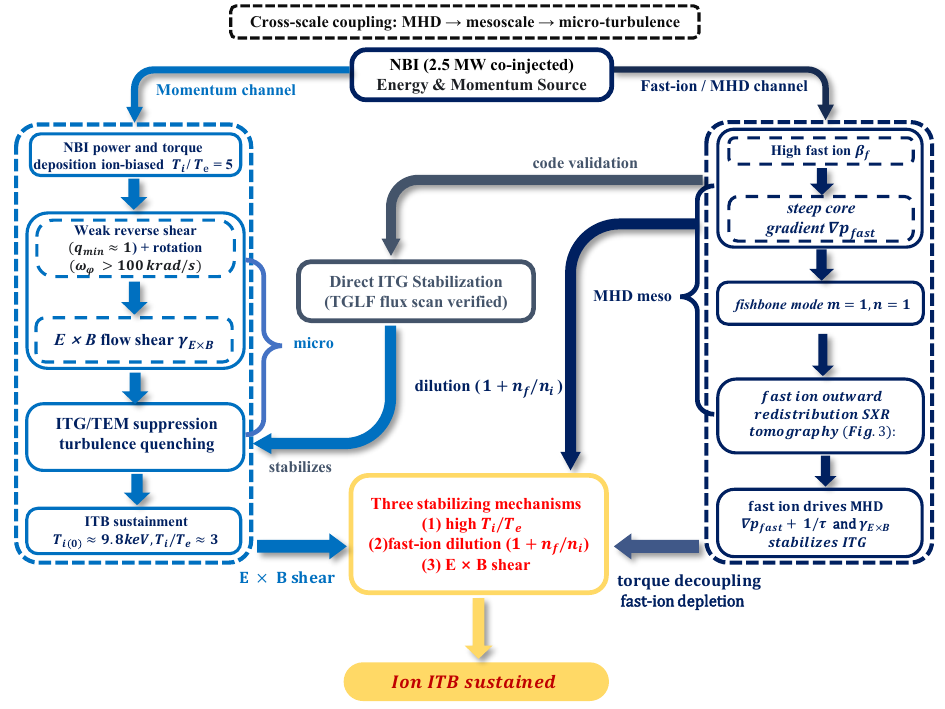}
 \caption{Cross-scale coupling mechanism spanning macroscopic MHD, fast-ion transport, and microturbulence. NBI drives two channels: the momentum channel (left) couples the NBI torque to toroidal rotation (above $100$\,krad/s) and $E\times B$ shear, suppressing ITG/TEM turbulence and sustaining the ITB. The fast-ion/MHD channel (right) excites an $m=1,n=1$ fishbone mode ($25$--$45$\,kHz) through the steep $\nabla p_{\mathrm{fast}}$ (peak $\beta_f\approx1.14\%$ at $t_3$) and drives outward fast-ion redistribution. The high temperature ratio $T_i/T_e$, fast-ion dilution, and $E\times B$ shear together sustain the ion ITB, and the fishbone reinforces them by stabilizing the ion-scale ITG and holding $q_{\min}$ near unity. The large-scale sawtooth, synchronized with the fishbone, temporarily weakens all three during the crash, before the post-crash recovery rebuilds the barrier.}
\label{fig19}
\end{figure}

Integrated OMFIT modeling with TGLF quasilinear simulations traces how one $m=1,n=1$ fishbone mode couples toroidal rotation, NBI torque, and ion-scale microturbulence in the HL-3 hot ion mode (shot \#11972, $T_i/T_e \approx 3$, $T_i(0) \approx 9.8$\,keV). Because the $45$\,keV deuterium beam lies below the critical energy $E_c \approx 62$\,keV, the NBI power couples mainly to the thermal ions and produces the extreme temperature ratio. The steep fast-ion pressure gradient built by 2.5\,MW NBI (peak $\beta_f \approx 1.14\%$) then both excites the $25$--$45$\,kHz fishbone and stabilizes the ion-scale ITG. On the momentum side the mode redistributes the torque-carrying fast ions outward, so the NBI torque enters the thermal channel further out and the torque source decouples from the core, while the fishbone holds $q_{\min}\approx 1$ to help sustain the ion internal transport barrier. The synchronized sawtooth crash then collapses the barrier and the rotation shear layer, dropping the toroidal rotation from above $100$ to $60$\,krad/s before the barrier rebuilds.
The SXR tomographic inversion (Fig.~\ref{fig3}) and the TGLF instability spectra (Fig.~\ref{fig10}) resolve two faces of one fast-ion population, bracketing the cross-scale coupling loop (Fig.~\ref{fig19}) that spans MHD, fast-ion transport, and microturbulence. Through resonant wave-particle interaction the fishbone continuously moves the resonant fast ions outward, and this redistribution is not a simple confinement loss. The fast ions that drive the mode also damp the ion-scale turbulence, stabilizing the trapped electron mode preferentially because they share its trapped phase space, while the thermal-ion ITG feels only the weaker dilution channel.

The loop closes in both directions. The outward shift empties the core phase space so the continuously injected beam ions deposit their power more readily, and the reduced anomalous transport lets the central $T_i(0)\approx 9.8$\,keV persist. In this beam-heated L-mode the fishbone therefore sustains, rather than erodes, confinement; the same mode in a burning plasma is instead a fast-ion loss channel, so the benign reading is specific to the absence of alpha particles and to injection that continuously replenishes the redistributed ions. Earlier studies followed the momentum and turbulence channels separately, so a single mode linking them was not resolved.

The ion ITB rests on three channels: a high temperature ratio inside the barrier, fast-ion dilution, and $E\times B$ shear at the barrier edge. Raising $T_i/T_e$ moves the high-wavenumber modes toward the electron diamagnetic direction, an ion temperature screening that damps the ion-direction ITG (Fig.~\ref{fig13}), and the fast ions amplify this suppression (Fig.~\ref{fig15}), so the fluxes of all species fall with $1/\tau$ (Fig.~\ref{fig12}). No single channel delivers this suppression on its own.
A steeper density gradient offers a further lever, distinct from the temperature ratio and the fast-ion population, and it acts in two opposite ways. As the drive of the trapped electron mode it sustains the low-wavenumber particle flux (Fig.~\ref{fig16}), yet through the $\eta$ parameter of Eq.~\eqref{eq:eta} it dilutes the temperature-gradient drive of the high-wavenumber branches. Because $a/L_{T_i}$ exceeds $a/L_{T_e}$ in the hot ion mode, $\eta_i > \eta_e$, so as $a/L_{n_e}$ rises the smaller $\eta_e$ drops below its instability threshold first and quenches the electron-direction branch while the ion-direction branch persists (Fig.~\ref{fig17}). This differential quenching, not any change in the underlying drive, turns the unstable branch from the electron to the ion direction and lowers the growth rate at fixed $T_i/T_e$.

In a burning plasma the alpha particles contribute to both MHD drive and microturbulence stabilization, so their density profile must balance the two. On HL-3 the weak magnetic shear leaves the NBI-torque-driven $E\times B$ shear as the main ITB sustainment mechanism. Toward a reactor, a higher triple product needs a high core temperature ratio $1/\tau$ with a steep $a/L_{T_i}$ inside the ITB, a combination the scans show crosses no high-transport region (Fig.~\ref{fig12}), and a strong ion edge barrier offers a complementary route.
A few limitations bound this analysis, it rests on a single discharge, so further HIM shots, with and without fishbone activity, would test the loop. The fluxes and growth rates are quasilinear (TGLF with the SAT2 rule), so their absolute levels are model-dependent and nonlinear gyrokinetic validation is deferred. The fast-ion redistribution is inferred from the SXR reconstruction and NUBEAM/ONETWO modeling rather than measured directly. The paper therefore offers a qualitative mechanism, not a calibrated predictive model of the fishbone--ITB coupling.

\ack{The authors thank General Atomics (GA) for the OMFIT integrated simulation platform, the HL-3 team for their support and the Southwestern Institute of Physics high-performance computing systems (Cluster02/03) for the computational resources used in this study.}

\funding{This work is supported by the National Key R\&D Program of China (Grant Nos 2025YFE03000200, 2024YFE03050004 and 2025YFE03000204), the National Natural Science Foundation of China (Grants Nos 12125502, 12535013, 12475216 and 12675277), the Nature Science Foundation of Sichuan Province (Grant No 2025ZNSFSC0833), the Innovation Program of Southwestern Institute of Physics (Grant No 202301XWCX001) and the Nuclear Technology Research and Development Program (Grant No HJSYF2024(02)).}


\data{The data that support the findings of this study are available from the corresponding author upon reasonable request.}

\suppdata{None.}

\section*{Appendix. OMFIT workflow}
\begin{figure}[htbp]
 \centering
        \includegraphics[width=0.6\textwidth]{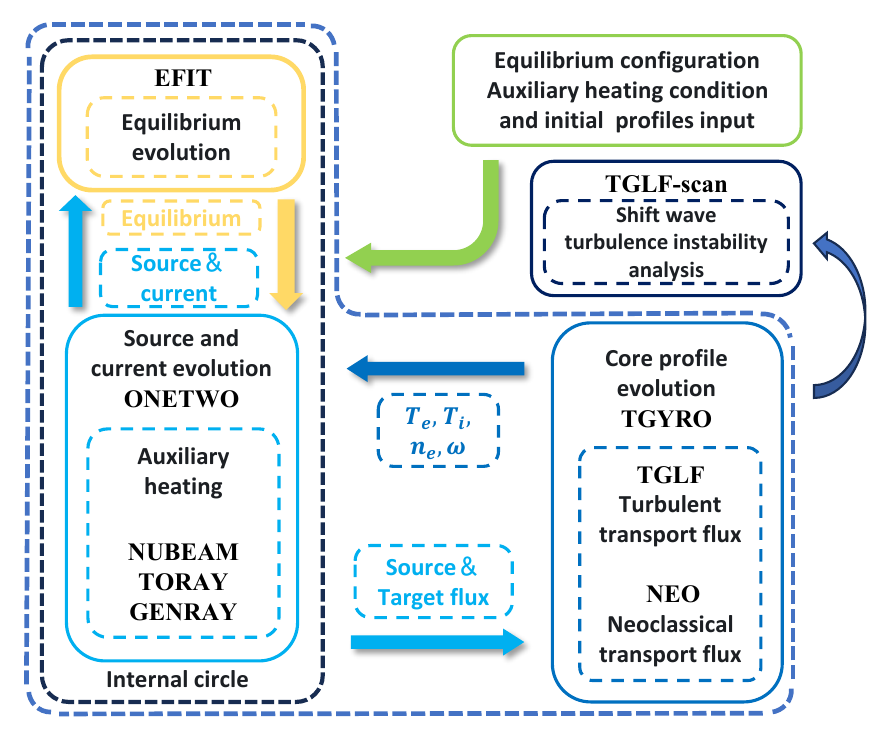}
 \caption{Schematic of the integrated OMFIT workflow. The equilibrium solver (EFIT), the source and current-drive module (ONETWO), and the transport module (TGYRO, used in flux-only mode) exchange profiles at each iteration until convergence. A companion TGLF-scan step then resolves the linear instability spectra and the flux sensitivity to selected profile quantities. Adapted from Ref.~\cite{shu2026integrated}.}
\label{fig20}
\end{figure}

Figure~\ref{fig20} collects the integrated modeling loop behind Sections~3--5, adapted from the companion study of shot \#13243~\cite{shu2026integrated}. In the present work the ONETWO source step is coupled to NUBEAM to track the deposition and torque of the two co-injected NBI beams. Each iteration passes the kinetic EFIT equilibrium to the source and current-drive calculation, which returns the heating and torque terms to the TGYRO transport solver; TGYRO is called in a flux-only mode to return the turbulent (TGLF, SAT2 rule) and neoclassical (NEO) transport coefficients and flux profiles, with no internal profile iteration, and the updated profiles are fed back into EFIT until the reconstructed zero-dimensional parameters settle against Table~\ref{tab1}. The converged profiles then seed the TGLF-scan instability and flux scans of Sections~4 and~5.

\bibliographystyle{iopart-num}
\bibliography{mybibfile}

\end{document}